\documentclass[twocolumn,final, a4paper, 10pt]{IEEEtran}
\usepackage{setspace}
\usepackage{latexsym}
\usepackage{graphicx}
\usepackage{comment}
\usepackage{amsfonts,amssymb,amsmath,bbm,url}
\usepackage[dvipsnames]{xcolor}
\def\BibTeX{{\rm B\kern-.05em{\sc i\kern-.025em b}\kern-.08em T\kern-.1667em\lower.7ex\hbox{E}\kern-.125emX}}
\newcommand{\ee}{{\rm e}}
\newcommand{\jj}{{\rm j}}  
\newcommand{\dd}{{\rm\,d}} 

\newcommand{\qv}{{\bf q}}

\newcommand{\uv}{{\bf u}}
\newcommand{\wv}{{\bf w}}
\newcommand{\vv}{{\bf v}}

\newcommand{\yv}{{\bf y}}
\newcommand{\zv}{{\bf z}}
\newcommand{\zerov}{{\bf 0}}

\newcommand{\Am}{{\bf A}}

\newcommand{\Cm}{{\bf C}}

\newcommand{\Gm}{{\bf G}}
\newcommand{\Hm}{{\bf H}}
\newcommand{\Id}{{\bf I}}

\newcommand{\Mm}{{\bf M}}

\newcommand{\Qm}{{\bf Q}}

\newcommand{\Tm}{{\bf T}}

\newcommand{\Wm}{{\bf W}}
\newcommand{\Vm}{{\bf V}}

\newcommand{\Zm}{{\bf Z}}

\newcommand{\Ac}{{\cal A}}
\newcommand{\Bc}{{\cal B}}

\newcommand{\Sc}{{\cal S}}

\newcommand{\gammav}{\boldsymbol{\gamma}}

\newcommand{\thetav}{\boldsymbol{\theta}}

\newcommand{\Thetam}{\boldsymbol{\Theta}}

\newcommand{\diag}{{\hbox{diag}}}

\def\trace{\mathsf{Tr}}

\def\Herm{^{\mathsf{H}}}
\def\Tran{^{\mathsf{T}}}

\def\ben{\begin{enumerate}}
\def\beq{\begin{equation}}
\def\beqa{\begin{eqnarray}}
\def\bit{\begin{itemize}}
\def\een{\end{enumerate}}
\def\eeq{\end{equation}}
\def\eeqa{\end{eqnarray}}
\def\eit{\end{itemize}}

\def\non{\nonumber\\}

\newtheorem{remark}{Remark}

\newtheorem{proposition}{Proposition}

\begin{document}

\title{IRS Configuration Techniques for Ultra Wideband Signals and THz Communications}

\author{Alberto Tarable, Laura Dossi,
 Giuseppe Virone,~\IEEEmembership{Senior Member, IEEE}, Alessandro~Nordio,~\IEEEmembership{Member, IEEE}
\thanks{A. Tarable, L. Dossi, G. Virone, A. Nordio are with
the National Research Council of Italy, Institute of Electronics, Information Engineering and Telecommunication  (CNR-IEIIT), 10129
Torino, Italy (e-mail: $<$name$>$.$<$surname$>$@ieiit.cnr.it).}
\thanks{This work was partially supported by the European Union
  under the Italian National Recovery and Resilience Plan (NRRP) of
  NextGenerationEU, partnership on ``Telecommunications of the
  Future'' (PE00000001 - program ``RESTART''), and by the Project:
  ``SoBigData.it - Strengthening the Italian RI for Social Mining and
  Big Data Analytics'' - Prot. IR0000013 - Avviso n. 3264 del
  28/12/2021. Part of this material was accepted for publication
  at ICC 2023 (Workshop) and can be found in~\cite{noiICC2023}.}
}
 
\maketitle
{\color{black}
\begin{abstract}
  Motivated by the challenges of future 6G communications where terahertz (THz)
  frequencies, intelligent reflective surfaces (IRSs) and
  ultra-wideband (UWB) signals coexist, we analyse and propose a set
  of efficient techniques for configuring the IRS when the signal
  bandwidth is a significant fraction of the central frequency (up to
  50\%).  To the best of our knowledge this is the first time that IRS
  configuration techniques are analyzed for such huge bandwidths.  In
  our work we take into account for the channel model,  the
  power spectral density of the signal reflected by the IRS and 
  the network geometry.  We evaluate the proposed solutions in terms of
  achievable rate and compare it against an upper bound we
  derived. Our results hint rules for designing IRS-aided
  communication systems and allow to draw conclusions on the trade-off
  between performance and complexity required for configuring the
  IRS.
\end{abstract}}


\section{INTRODUCTION\label{sec:intro}} 
{\color{black} THz communications
 is a promising technology able to fulfill the ambitious goals in
 terms of capacity of future 6G wireless networks, which are expected to
 reach Tb/s data rates, and handle UWB
 signals~\cite{Huq2019}. Future implementation of THz communications
 need to face harsh propagation environments typical of
 sub-millimeter wavelengths and characterized by high path losses and
 blockages.
While path loss can be compensated for by using high-gain antenna
arrays and through beamforming~\cite{DRL-beamforming}, the
availability of LoS links is crucial for THz communications. In their
absence, blockages can be circumvented e.g. by
deploying
IRSs,
which are one of the key elements envisioned for building smart radio
environments (SRE)~\cite{DiRenzo,HMIMO}.

The interaction between UWB signals
with large antenna arrays and IRSs poses, hovever, a number of
new challenges. As an example, the signal propagation delay across
large arrays has to be taken into account. Indeed, if the number of
antenna elements in the array is large, the propagation delay becomes
a significant fraction of the symbol period. Such a phenomenon, 
known as spatial-wideband effect~\cite{SpatialWideband} in the literature on array and
radar signal processing, can entail performance
losses. Its frequency-domain manifestation is often referred to as {\em beam squint}%
~\cite{Delay_phase_precoding2022,WB_beamforming_mMIMO}.
As a typical scenario, beams generated by antenna arrays or IRSs with
frequency-independent adjustable phase shifters do not point to the
same direction for different frequencies. As a consequence, only some
frequencies can take advantage from high array gain toward the desired
direction.
This effect becomes more prominent as both the size of the antenna array
and the signal bandwidth increase. Without taking
countermeasures, an IRS-aided communication channel, as that depicted
in Figure~\ref{fig:network}, is prone to beam-squint effects
introduced by both the transmitter antenna array and the IRS. To solve
these problems, several solutions to be applied at the transmitter
have been proposed in order to make the generated beams nearly
frequency-insensitive.  Instead, in this work we propose to
  operate on the IRS by properly configuring its elements so as to
  compensate for the beam-squint effect, in the presence of
  UWB signals. In the following we provide an overview of
  the contributions of our work, followed by a review
  of the state of the art.

\subsection{Contributions of this work}
Motivated by the challenges of future 6G communications and by the
recent advances in IRS technology~\cite{Behrooz2023}, in this work
\begin{itemize}
 \item we study a communication system where
   the signal bandwidth is a significant fraction (i.e., up to 50\%)
   of the central frequency. To the best of our knowledge,  this is the
   first time that such huge bandwidth is considered in IRS-aided
   communication;
 \item we consider a set of effective techniques for configuring the
   IRS phase-shifts; although some of these have already been proposed
   for narrowband communications, we have adapted them to signal bandwidths
   and power spectral densities that can be envisioned for 6G communications;
\item we provide a set of numerical results assessing the performance
  of the above mentioned techniques, which hint design rules of IRS-aided
  communication systems and allow to draw conclusions on the trade-off
  between performance and complexity;
\item we derive an upper bound to the achievable rate for a system,
  working in the THz frequencies, composed of a base station (BS), an
  IRS and an intended receiver (UE); such bound is used as a benchmark
  for assessing the performance of the considered IRS configuration
  techniques.; by numerical analysis, we prove the bound to be fairly
  tight, for a wide range of signal bandwidths;
\item finally, we provide an analytical condition granting the
  optimality of an IRS configuration technique, as a function of the
  signal bandwidth, of the delay spread, and of the system geometry.
\end{itemize}

Summarizing, this work, which is an extension of~\cite{noiICC2023},
aims at understanding how a THz communication system should be
designed to take advantage of IRSs in the presence of UWB signals. It
provides fundamental performance limits for IRS-assisted systems in a
realistic setting and proposes practical strategies
that approach quite closely such performance limits.
  
  \subsection{Related works}
 Recent research in IRS-aided communications encompasses contributions
 in IRS modeling, wideband channel estimation and modeling, and
 mitigation of the beam-squint effect.

 In many works the response of IRS elements is modeled as
 frequency-independent in both phase and amplitude
 although, in practice, it is far from being
 ideal~\cite{survey, wideband_IRSmodel,Zeng2022}.  The assumption of
 frequency independence of the IRS response is a good approximation
 only in the IRS operational bandwidth, which typically amounts to a
 small fraction of the operational central frequency. 
 For this reason, so far, IRS did not seem suited to support UWB signals, which are
 characterized by relative bandwiths greater than 20\%.

 However, a very recent research in IRS design, has proposed an IRS
 model operating at 26.5\,GHz~\cite{Behrooz2023} showing an
 operational bandwidth of about $30\%$ of the carrier frequency, thus
 enabling IRS-aided UWB communications. Imagining further technology
 breakthrough, we expect that such figures could be further raised in
 the future and that very soon IRS-aided UWB communications will 
 be available at much higher frequencies, e.g., in the sub-THz range.

The problem of optimally configuring the IRS in the presence of
wideband signals has been considered in a few very recent papers.  The
authors in~\cite{BeamSquintMitigating} consider a multicarrier
wideband signal and propose to configure the IRS so as to maximize the
achievable rate, for orthogonal frequency-division multiplexing
signals, central frequency 28 GHz and relative bandwidth of
about 7\%, much smaller than that considered in this work.
Similarly to what we do,
they also propose a technique based on singular-value decomposition of
a channel matrix. However, they apply such technique to the IRS--UE
channel only, and neglect the influence of the BS--IRS channel and of
the signal power spectral density. 

Authors in~\cite{wideband_THz} introduce a generalized channel model
suitable for large IRSs operating in near field at THz frequencies,
while~\cite{rainbow,UWB_THz_IRS} propose to exploit the beam-squint
effect, rather than eliminating it, to generate multiple beams
simultaneously focused on multiple targets. Finally,
\cite{WB_estimation} accounts for the beam-squint effect while
addressing the problem of estimating the wideband channel in far
field.

  Beam-squint mitigation at the antenna arrays is considered in many
  works and solved  e.g. through delay-phase precoding techniques~\cite{Delay_phase_precoding2022}
  or true-time delayers (TTDs), in  order to achieve ultra-wideband beamforming. The work
  in~\cite{Hashemi2008} discusses the implementation of TTDs and their
  applications for signals having instantaneous ultra-wide bandwidths,
  while in~\cite{WB_beamforming_mMIMO} approaches based on virtual
  sub-arrays and TTD lines are considered for eliminating beam squint.
  Such architecture can achieve performance very close to fully digital
  transceivers, although requiring high hardware cost and
  large power consumption.
Unfortunately, these are not viable methods for eliminating beam squint at the 
IRSs, since they are made of passive elements, whose phase
shifts cannot be controlled in the frequency domain.
The recent proposal in~\cite{DelayAdjustable_Hanzo} consists in a
new implementation of the IRS, called delay-adjustable metasurface
(DAM), where the elements rely on varactor diodes and are able to impose
a controllable extra delay onto the reflected signals. DAMs are,
however, subject to power losses and in a very early concept stage, as
opposed to standard IRSs, whose prototypes are already available, as reported for
example in~\cite{prototype}.

The remainder of the paper is organized as follows.
Section~\ref{sec:model} describes the system model and formulates the
optimization problem.  Section~\ref{sec:UB} derives an upper bound to
the achievable rate, Section~\ref{sec:max_Pr} proposes several
algorithms aiming at maximizing the received power, whose performance
are then assessed and compared in
Section~\ref{sec:results}. Conclusions are drawn in
Section~\ref{sec:conclusions}.
} 
\subsection{Mathematical notation}
Boldface uppercase and lowercase letters denote matrices and vectors,
respectively, uppercase calligraphic letters denote sets.
$\Id_k$ is the $k \times k$ identity matrix.  The $(i,j)$-th element
of matrix $\Am$ is denoted by $[\Am]_{i,j}$,  $\Am\Herm$ is its conjugate transpose
 and $\trace\{\Am\}$ its trace.  We denote by $\rho(\Am)$ the  rank of  matrix
$\Am$.  The symbol $\odot$ denotes the Hadamard (i.e.,  elementwise) product 
Finally,  $\mu(\Ac)$ denotes the measure of the set $\Ac$.

\section{System model and problem formulation\label{sec:model}}

\begin{figure}[t]
  \centering
\includegraphics[width=0.9\columnwidth]{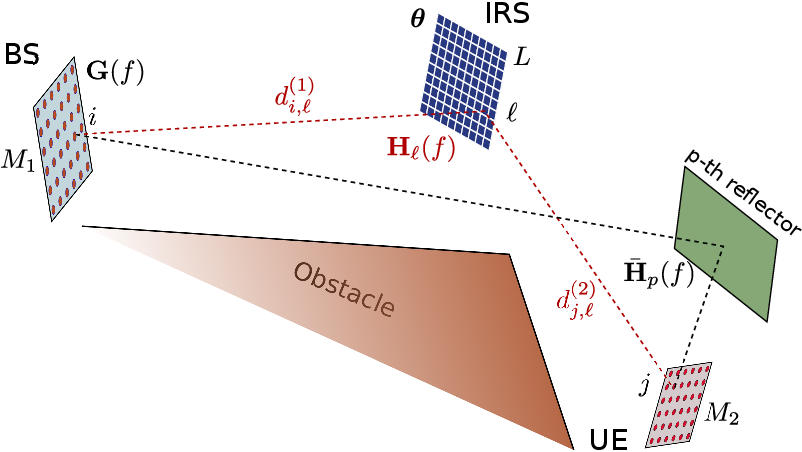}
\caption{Representation of a communication network where a BS and a UE
  communicate by exploiting an IRS. The LoS link between BS and UE is
  unavailable due to an obstacle. Channel multipath components are due
  to the presence of a small number of reflectors.}
\label{fig:network}
\end{figure} 
We consider a wireless system operating in the THz band where a base
station (BS), equipped with an array of $M_1$ antennas, transmits a data stream to
a user (UE), equipped with $M_2$,antennas,  as depicted in
Fig.~\ref{fig:network}. The transmitted signal has bandwidth $B_w$ and is characterized by the
$M_1\times M_1$ power spectral density matrix $\Gm(f)$,
which we assume to be positive semidefinite and whose support,
in the frequency space, is contained in set $\Bc$. The BS transmit power is
denoted by
\begin{equation}
  P^{\rm tx} = \int_{\Bc} \trace\{\Gm(f)\} \dd f\,. \label{Eq:Pt}
\end{equation}
At sub-THz and THz
frequencies
scattering and diffraction effects provide a marginal contribution to
propagation.  Indeed, most of the received energy is due to the LoS component and,
possibly, to some non-LoS (NLoS) reflected rays~\cite{noiTWC}.
Moreover, at THz frequencies many other
effects come into play, such as molecular absorption~\cite{Molecular}, blockage, and
large-scale fading effects (shadowing).  In our model, we suppose that the
UE is not in LoS with the BS due to blockage, while
some rays leaving the BS may reach the UE after being reflected on 
large surfaces (e.g., building walls).  In order to cope with the
lack of LoS connectivity, the system exploits an IRS, composed of $L$
electronically configurable reflective elements, as shown in Fig.~\ref{fig:network}.

\subsection{IRS model}

Practical IRSs are designed to work in a specific frequency band
$\Bc$, around a central frequency $f_c$. Such operational band can be
defined as the range of frequencies around $f_c$ in which the transfer
function of the $\ell$-th IRS element, $\ell = 1,\dots,L$, satisfies
  \begin{equation}
    \zeta_{\ell}(f)  \simeq  \widetilde{\zeta}(f)\ee^{\jj \theta_{\ell}}, \quad \forall f\in \Bc
  \label{eq:zeta_ell_approx}
  \end{equation}
  where $\theta_\ell$ is a supposedly controllable phase shift and
  $\widetilde{\zeta}(f)$ is a {\color{black} known} non-controllable complex frequency
  response, independent of $\theta_\ell$, with
  $|\widetilde{\zeta}(f)|$ constant for $f\in \Bc$.  In practice, the
  deviation of the true frequency response
  from~\eqref{eq:zeta_ell_approx}, in both magnitude and phase, should
  have a limited effect on performance.  Many works in the literature
  (see e.g.~\cite{survey} and referencese therein) assume that, in the
  operational band, IRS elements are ideal phase shifters, i.e., with
  $\widetilde{\zeta}(f) = 1$.

  At present, most IRS
  implementations work in the sub-6\,GHz band, whereas few are
  reported to operate in the mm-wave bands.
  For these
  implementations, the operational bandwidth greatly varies and
  usually amounts to a small fraction of $f_c$, typically less than
  $10\%$.  However, an IRS design operating at
  $f_c=26.5$\,GHz and allowing an operational bandwidth of up to 30\%
  was recently proposed~\cite{Behrooz2023}.  This pioneering result enables
  IRS-assisted UWB communications in the mm-wave bands, making it a
  good candidate for 5G applications.
  Imagining further breakthroughs that, similarly
  to~\cite{Behrooz2023}, could extend to THz frequencies the
  operational band of IRSs, maybe with on-chip technology, this paper
  assumes that the in-band frequency response of IRS elements
  satisfies~\eqref{eq:zeta_ell_approx}. Even more generally, we will
  not need to pose any constraint on $|\widetilde{\zeta}(f)|$, which
  can then be an arbitrarily-varying frequency function. Instead, it
  is important to stress that $\widetilde{\zeta}(f)$
  in~\eqref{eq:zeta_ell_approx} is independent of $\theta_\ell$ in our model.
  
\subsection{Overall system model}
The channel transfer function of the BS--UE link is represented
by the $M_1\times M_2$ matrix $\Hm(f,\thetav)$ given by
\begin{eqnarray}
  \Hm(f, \thetav)
  &=& \sum_{\ell=1}^L \ee^{\jj \theta_\ell} \Hm_{\ell}(f) + \bar{\Hm}(f)\label{eq:H_f} 
\end{eqnarray} 
where $\thetav=[\theta_1,\ldots, \theta_L]\Tran$, $\Hm_{\ell}(f)$ is
the channel transfer function for the path through the $\ell$-th IRS
element, which accounts for both IRS frequency response and signal
propagation.  The matrix $\bar{\Hm}(f)$
takes into account the contribution of multipath, i.e, in this case, of rays reflected by large
obstacles.  Note that in~\eqref{eq:H_f} we stress the dependence of
the channel matrix on the IRS configuration, $\thetav$, by explicitly
indicating it as an argument.
In details, $[\Hm_{\ell}(f)]_{i,j}$ is the scalar transfer function
corresponding to the path connecting the $i$-th BS antenna to
the $j$-th UE antenna through the $\ell$-th IRS element, whose length is $d_{i,j,\ell}$(see
Figure~\ref{fig:network}),  and can be modeled as
\begin{equation}\label{eq:Hl_f}
 \hspace{-1ex}[\Hm_{\ell}(f)]_{i,j} {=}
  \alpha^{(1)}_{i,\ell} \alpha^{(2)}_{j,\ell} \widetilde{\zeta}(f)\ee^{-\jj 2\pi f \tau_{i,j,\ell}-\kappa(f)
    d_{i,j,\ell}}\frac{ \sqrt{A^{(1)}_{i,\ell}
  A^{(2)}_{j,\ell}}}{4\pi
    d^{(1)}_{i,\ell} d^{(2)}_{j,\ell}}
\end{equation} 
where $\tau_{i,j,\ell}{=}\frac{d_{i,j,\ell}}{c}$ 
is the total travel time of the path.  In particular,
$d_{i,j,\ell} {=} d^{(1)}_{i,\ell}{+} d^{(2)}_{j, \ell}$ where
$d^{(1)}_{i,\ell}$ ($d^{(2)}_{j,\ell}$) is the distance between
transmit antenna $i$ (receive antenna $j$) and IRS element $\ell$.
The superscripts $(1)$ and $(2)$ refer  to the BS--IRS and
IRS--UE hops,  respectively.  With the same notation $\alpha^{(1)}_{i,\ell}$ and
$\alpha^{(2)}_{j,\ell}$ take into account the
antenna gains at the BS and UE,  respectively,  and possible shadowing effects, while
$A^{(1)}_{i,\ell}=A\cos(\phi^{(1)}_{i,\ell})$ and $A^{(2)}_{j,\ell}=A\cos(\phi^{(2)}_{j,\ell})$ are the effective areas of
the IRS element $\ell$ as observed, respectively, from the $i$-th BS
antenna and the $j$-th UE antenna~\cite{noiTWC},  where $A$ is the
physical area of the IRS meta-atom and $\phi^{(1)}_{i,\ell}$
($\phi^{(2)}_{j,\ell}$) is the angle of arrival from the $i$-th BS
antenna (angle of departure towards the $j$-th UE antenna), measured counterclockwise
w.r.t. the normal to the $\ell$-th IRS meta-atom.  Finally,
$\kappa(f)$ is the
molecular-absorption coefficient~\cite{propagation}.
The multipath matrix $\bar{\Hm}(f)$ can be written as
\begin{equation}
  \bar{\Hm}(f) = \sum_{p=1}^P\bar{\Hm}_p(f) \label{eq:barH}
\end{equation}
where $P$ is the number of rays and $\bar{\Hm}_p(f)$ is the channel matrix corresponding to the $p$-th reflected ray. According to the image theorem we have
\[ [\bar{\Hm}_p(f)]_{i,j} = \rho_p\alpha_{p,i,j}\frac{c}{4\pi f
    d_{p,i,j}} \ee^{-\jj 2\pi f \tau_{p,i,j}}\ee^{-\kappa(f)
    d_{p,i,j}} \]
where $\rho_p$ is the reflection coefficient of the $p$-th reflector, $d_{p,i,j}$ is the length of the path connecting the $i$-th
  BS antenna  to the $j$-th UE antenna through the
  $p$-th reflector, and $\tau_{p,i,j}= \frac{d_{p,i,j}}{c}$ is the
  corresponding travel time. Finally $\alpha_{p,i,j}$ accounts for
  possible shadowing.
  
  \begin{remark}
    Note that the model in~\eqref{eq:zeta_ell_approx}, ~\eqref{eq:H_f}
    and~\eqref{eq:Hl_f} is very general and can apply to a 3D
    geometric environment with IRS and antenna arrays of any shape
    Moreover, many results in the
    following, such as those of Section~\ref{sec:UB}, hold for any
    channel model satisfying~\eqref{eq:H_f}, regardless of the
    particular expression of $\Hm_{\ell}(f)$ and $\bar{\Hm}(f)$.
\end{remark}

\subsection{Rate optimization problem}

Given the power spectral density matrix of the input,
$\Gm(f)$, the achievable rate is
\begin{eqnarray}
  \hspace{-2ex}R(\thetav)
  &\hspace{-2ex}=& \hspace{-2ex}\int_{\Bc} \log \left|\Id \mathord{+} \frac{\Hm(f,\thetav) \Gm(f) \Hm\Herm(f,\thetav)}{N_0}\right| \dd f\,, \label{eq:R}
\end{eqnarray}
where $N_0/2$ is the per-dimension power spectral density of the
circularly-symmetric additive white Gaussian noise at the receiver.
Our goal is to optimize the IRS phase shifts, $\thetav$, so as to
maximize the rate $R(\thetav)$. In other words, we would like to solve:
\begin{equation} \max_{\thetav}
  R(\thetav)\,. \label{eq:problem}\end{equation}
  This  maximization is difficult to tackle; indeed, an explicit closed-form expression for the
  maximizer does not exist, so that one must resort to  heuristic
  approaches in order to find good, albeit suboptimal, solutions. To that end, we start by
  deriving an upper bound to~\eqref{eq:R}, hinting a solution for~\eqref{eq:problem},
  which will be shown to be relatively tight in Section~\ref{sec:results}.
  Such bound will be used as a benchmark to assess the performance of
  suboptimal solutions to~\eqref{eq:problem}.
  
\section{Upper bound to the end-to-end communication rate\label{sec:UB}} In order to derive the
  upper bound to~\eqref{eq:R}, we first give a couple of
  definitions. Let 
\begin{equation}\label{eq:P_r}
  P^{\rm rx}(\thetav) = \int_{\Bc}\trace\{\Hm(f,\thetav) \Gm(f)\Hm\Herm(f,\thetav)\} \dd f
\end{equation}
be the received power and, let $\Tm$ be the $L \times L$ matrix
whose $(\ell,\ell')$ entry is given by
\begin{equation}
  [\Tm]_{\ell,\ell'} =  \int_{\Bc}\trace\{\Hm_{\ell}(f)\Gm(f)\Hm_{\ell'}\Herm(f)\}\dd f\,.\label{eq:Tm} 
  \end{equation}
Then, we can state the following proposition:
\begin{proposition}\label{prop:1}  
For $M_1\ge M_2$ and $L \ge M_2$, the maximum rate can be upper-bounded as
\begin{eqnarray}
  &&\hspace{-3ex}\max_{\thetav} R(\thetav)\non
  &\hspace{-5ex}\le&\hspace{-4ex}  \sum_{m=1}^{M_2}B_m\log\left(1 + \frac{\max_{\thetav} P^{\rm rx}(\thetav)}{N_0\sum_{i=1}^{M_2}B_i}\right)   \label{eq:upper_bound} \\
  &\hspace{-5ex}=& \hspace{-4ex} \sum_{m=1}^{M_2}B_m\log\left(1\mathord{+}\frac{\displaystyle\max_{\gammav, |\gamma_{\ell}|=1, \forall \ell}\gammav\Herm \Tm \gammav \mathord{+}w \mathord{+}2\Re\left\{\qv\Herm \gammav\right\}}{N_0\sum_{i=1}^{M_2}B_i}\right)   \label{eq:upper_bound2} \\
  &\hspace{-5ex}\le&\hspace{-4ex}  \sum_{m=1}^{M_2}B_m\log\left(1 + \frac{L\lambda_T^{\max}+w+2\sum_{\ell=1}^L|q_{\ell}|}{N_0\sum_{i=1}^{M_2}B_i}\right)\, \label{eq:upper_bound3}
\end{eqnarray}
where $N_0$ is the noise power
spectral density, $\gammav=[\gamma_1,\ldots,\gamma_L]\Tran$, $\gamma_{\ell}=\ee^{\jj \theta_{\ell}}$, for $\ell=1,\ldots,L$. The coefficients $B_m$ are defined as
\beq \label{eq:B_m}
B_m = \int_{\Bc} \mathbbm{1} \{ r_Q(f) \geq m\} \dd f, \,\,\, m = 1,\dots,M_2\ \,,
\eeq
being $r_Q(f)$ any upper bound to the rank of the matrix $\Hm(f,\thetav) \Gm(f) \Hm\Herm(f,\thetav)$ for all $\thetav$. Moreover 
$\lambda_T^{\max}$ is the maximum eigenvalue of
$\Tm$,  the scalar $w$ is given by $w=\int_{\Bc}\trace\{\bar{\Hm}(f)\Gm(f)\bar{\Hm}\Herm(f)\}\dd f$
and $\qv$ is a length-$L$ vector with entries 
\beq \label{eq:q_ell}
q_\ell^* =  \int_{\Bc}\trace\{\Hm_{\ell}(f)\Gm(f)\bar{\Hm}\Herm(f)\}\dd f,\,\,\,\ell=1,\ldots,L\,.
\eeq
\end{proposition}

In the special case where $\Gm(f)$ is rank-1 on its support, the bound~\eqref{eq:upper_bound2} reduces to 
\begin{equation}
  \max_{\thetav} R(\thetav)  \le B_1\log\left(1 + \frac{\max_{\thetav} P^{\rm rx}(\thetav)}{N_0B_1}\right)\,.
  \label{eq:upper_bound_rank1}
\end{equation}

The above proposition holds for a general channel
satisfying~\eqref{eq:H_f}. In particular, if~\eqref{eq:Hl_f} is also
satisfied, we can give an explicit expression for $r_Q(f)$, as per the
following proposition.
\begin{proposition}\label{prop:1bis} 
For a channel model satisfying \eqref{eq:H_f}-\eqref{eq:Hl_f}, the upper bound $r_Q(f)$
can be written as 
\begin{equation}
  r_Q(f) =  \min\left\{M_2, \rho(\Gm(f)), \min\{L,\rho(\Vm(f)),\rho_{wh}(f))\right\}\,. \label{eq:rank_bound}
\end{equation}
where $\rho_{wh}(f) \triangleq \rho(\Wm(f))\} + \rho(\bar{\Hm}(f)$, the matrix $\Hm_{\ell}(f)$ can be decomposed as
$\Hm_{\ell}(f)=\vv_{\ell}(f)\wv_{\ell}\Tran(f)$ and the matrices
$\Vm(f)$ and $\Wm(f)$ are given by $\Vm(f) = [\vv_1(f),\ldots,
  \vv_L(f)]$ and $\Wm(f) = [\wv_1(f),\ldots, \wv_L(f)]$, respectively.  Note that, in
practice, $L\gg M_2$, so the term $L$ can be removed from the above
expression.
\end{proposition}

 The proofs of both propositions are provided in~\ref{app:prop:1}.

\begin{remark}
  Note that in the low-SNR regime, i.e., with received SNR
  $\eta=\frac{P^{\rm rx}(\thetav)}{B_1N_0}\ll 1$, the rate $R(\thetav)$
  in~\eqref{eq:R} can be approximated to the first order as
  $R(\thetav)= B_1 \log(1 +\eta) + O(\eta^2)$, and the maximization over $\thetav$ leads to~\eqref{eq:upper_bound_rank1}.
\end{remark}
 
\section{Sub-optimal IRS configuration techniques\label{sec:max_Pr}}
  As already said, an
  explicit expression for the maximizer of~\eqref{eq:problem} is
  difficult to obtain. However, we observe that the bound in~\eqref{eq:upper_bound}
  increases as the received power, $P^{\rm rx}(\thetav)$, increases.
  We then propose to solve
  \begin{equation}
  \max_{\thetav}P^{\rm rx}(\thetav) = \max_{\gammav, |\gamma_{\ell}|=1, \forall \ell}\left(\gammav\Herm \Tm \gammav +w +2\Re\left\{\qv\Herm \gammav\right\}\right)\label{eq:UCQP}\,,
\end{equation}
i.e., to maximize $P^{\rm rx}(\thetav)$, which corresponds to
maximizing the upper bound to the rate, instead of the actual
rate. Now,~\eqref{eq:UCQP} is an example of Unimodular-Constraint
Quadratic Problem (UCQP), which is known to be NP-hard.
{\color{black} Note that a similar problem has been faced in~\cite{Pan_2020} in the context of simultaneous IRS-aided
  communication and power transfer, and solved using a successive convex approximation technique.}
 To solve~\eqref{eq:UCQP} we propose two strategies:
  \begin{itemize}
  \item employing a numerical algorithms designed for solving UCQP
    problems, as described in Section~\ref{sec:numeric_alg}; 
  \item applying a heuristic solution based on the eigenvalue
    decomposition of matrix $\Tm$, as described in
    Section~\ref{sec:heuristic}, which is proved to be high-performance.
  \end{itemize}
  
  Additionally, in Section~\ref{sec:NB_solutions}, we propose a set of
  ``narrowband solutions'' to~\eqref{eq:problem}. Some of them show
  interesting performance under some conditions on the system
  parameters and geometry, as described in detail in
  Section~\ref{sec:results}. 

\subsection{Numerical algorithms solving UCQP problems\label{sec:numeric_alg}}
  There are several numerical algorithms able to
  provide close-to-optimal solutions to UCQP problems, such as
  MERIT~\cite{Stoica14} and SCF~\cite{Aldayel17}. Among them, we consider SCF, which takes as input
  the $L\times L$ matrix $\Tm$ and applies an iterative procedure: at
  each iteration, it computes the inverse of a $2L\times 2L$ real
  matrix.  Although it can provide near-optimum solution
  to~\eqref{eq:UCQP}, its complexity rapidly increases with $L$ and
  also increases with the number of iterations required to reach
  convergence. The algorithm reported in~\cite{Aldayel17} has been
  adapted to the problem at hand. Its performance is reported in
  Section~\ref{sec:results} and compared against other techniques.

\subsection{A heuristic solution based on eigenvalue
    decomposition\label{sec:heuristic}}
An heuristic solution for the maximizer of~\eqref{eq:UCQP} can be
derived in the absence of channel multipath components
($\bar{\Hm}(f)=\zerov$). In order to obtain it, we proceed as follows:
\begin{itemize}
\item we first relax the UCQP problem to
  \begin{equation} \max_{\gammav, \gammav\Herm\gammav=L}\gammav\Herm \Tm \gammav\label{eq:UCQP_no_mpath}\end{equation}
   and compute its maximizer $\widetilde{\gammav} = \sqrt{L}\uv^{\max}$  where $\uv^{\max}$
   is the unit-norm eigenvector of $\Tm$ corresponding to its maximum
   eigenvalue. Clearly $\widetilde{\gammav}$ does not, in general,
   satisfy the unimodular constraint $|\gamma_{\ell}|=1, \forall \ell$;
 \item we then extract the phases of each entry of $\widetilde{\gammav}$, obtaining 
\begin{equation}\label{eq:eig}
  \gamma_{\ell}^\star = \ee^{\jj \arg(\widetilde{\gamma}_\ell)},\qquad \ell=1,\ldots,L \,.
\end{equation}
and we use 
  $\gammav^\star=[ \gamma_1^\star,\ldots, \gamma_L^\star]\Tran$ as an approximate
solution for the original problem~\eqref{eq:UCQP}.
\end{itemize}
A similar procedure was also suggested in~\cite{BeamSquintMitigating}
for a NLoS  scenario where, however, it was applied to the IRS--UE
channel only. Instead, we apply it to matrix $\Tm$ which includes
the effect of both the TX-RX channel matrix and the signal power
spectral density. This approach requires the knowledge of 
$\Tm$ and its complexity is dominated by the extraction of the largest eigenvalue of $\Tm$.

\subsection{A family of solutions inspired by narrowband signals\label{sec:NB_solutions}} 

In this subsection, we consider a set of suboptimal
  solutions to the problem~\eqref{eq:UCQP}, tailored to the case where
  the transmitted signal is narrowband at frequency $f_0$ and the communication channel
  does not exhibit multipath components (i.e., $\bar{\Hm}(f)=0$).
  Such solutions take, in general, the form
\begin{equation}\label{eq:NB}
\gamma^{\rm NB}_{\ell}(f_0) = \ee^{\jj 2\pi f_0 \tau_{\ell}}, \qquad \ell = 1,\ldots,L\,.
\end{equation}
They satisfy the unimodular constraint, and will be called {\em
  narrowband (NB) at frequency $f_0$ with coefficients $\tau_{\ell}$}.
They are interesting in that they show in some cases good performance,
even in the presence of wideband signals. Narrowband
  solutions of the family~\eqref{eq:NB} have already been proposed
  in~\cite{BeamSquintMitigating,Larsson_2020}, however they have been
  investigated only for small signal bandwidths, compared to the central frequency $f_c$.

In Section~\ref{sec:results},
we will compare the rate they achieve against the upper bound
in~\eqref{eq:upper_bound3}, for a wide variety of system parameters and
geometries.

The narrowband solutions we propose are derived under the hypothesis
of small BS and UE array size, and consider also the case of discrete
signal spectrum, as detailed in the following
Sections~\ref{sec:small_arrays} and~\ref{sec:discrete_spectrum},
respectively. In Section \ref{sec:NB_opt} we also describe the optimal
NB solution maximizing~\eqref{eq:UCQP}.

\subsubsection{Small BS and UE array size\label{sec:small_arrays}}
We here assume that the BS and UE antenna arrays are small w.r.t. the BS--IRS and
IRS--UE distances, i.e., they can be seen under a small angle when
observed from the IRS. Then, we can decouple the indices
$i,j,\ell$ in the expression for $\tau_{i,j,\ell}$ in~\eqref{eq:Hl_f}  and write it as
a sum of three terms as
\[ \tau_{i,j,\ell} \approx c_{{\rm BS},i} + c_{{\rm UE},j}+
  \tau_{\ell}\] where $c_{{\rm BS},i}$ and $c_{{\rm UE},j}$ are functions of the geometrical arrangement of the array
elements at the BS and at the UE,
respectively, as well as of the array orientation
in space w.r.t. the IRS. The term $\tau_{\ell}$ represents the signal
travel time through IRS element $\ell$, measured from the center of
the BS array to the center of the UE array.  Under the same
assumption, and referring to~\eqref{eq:Hl_f}, we can approximate the
distances $d^{(1)}_{i,\ell}$ and $d^{(2)}_{j,\ell}$ as
$d^{(1)}_{i,\ell} \approx d^{(1)}_{\ell}$ and
$d^{(2)}_{j,\ell} \approx d^{(2)}_{\ell}$ so that
$d_{i,j,\ell}\approx d_{\ell} = d^{(1)}_{\ell}+d^{(2)}_{\ell}$.
Moreover, also the antenna gains and the shadowing coefficients do not
depend on $i$ and $j$, i.e.,
$\alpha^{(1)}_{i,\ell}=\alpha^{(1)}_{\ell}$,
$\alpha^{(2)}_{j,\ell}=\alpha^{(2)}_{\ell}$ and the dependence on $i$
and $j$ can also be dropped from $A^{(1)}_{i,\ell}=A^{(1)}_{\ell}$
and $A^{(2)}_{j,\ell}=A_{\ell}^{(2)}$, since the IRS elements have the
same area and see all BS (UE) antenna elements under the same angle.
Finally, if the attenuation due to molecular absorption is
  negligible for every $f\in \Bc$ (as for indoor application and
  short-range communications), the term $\ee^{\kappa(f)d_{i,j,\ell}}$
  in ~\eqref{eq:Hl_f} can be neglected.  
By
plugging the above approximations in~\eqref{eq:Hl_f}, the channel
matrix $\Hm_{\ell}(f)$ can be simplified as
\begin{equation} \label{eq:hyp1}
  \Hm_{\ell}(f) = K_{\ell}\Am(f)\ee^{-\jj 2\pi f \tau_{\ell}}
\end{equation}
where we defined 
$[\Am(f)]_{i,j}\mathord{=} \widetilde{\zeta}(f)\ee^{-\jj 2\pi f
  (c_{{\rm BS},j} + c_{{\rm UE},i})}$ and
$K_{\ell}\mathord{=}\frac{\alpha^{(1)}_{\ell}\alpha^{(2)}_{\ell}\sqrt{A^{(1)}_{\ell}A^{(2)}_{\ell}}}{4\pi
  d^{(1)}_{\ell}d^{(2)}_{\ell}}$.
For this scenario, we have the following propositions:

\begin{proposition} \label{prop:2}
If the channel matrix $\Hm_{\ell}(f)$ takes the form
in~\eqref{eq:hyp1} and the signal is monochromatic at frequency $f_0$,
the solution to the problem~\eqref{eq:UCQP_no_mpath} is~\eqref{eq:NB}.
\end{proposition}

The proof is provided in~\ref{app:prop:2}.
Such result can be generalized to wideband signals as follows.

\begin{proposition} \label{prop:3}
Consider a wideband signal and a channel for which \eqref{eq:hyp1}
holds. Let $T_{\rm DS} = \max_{\ell, \ell' = 1,\dots,L} (\tau_{\ell} -
\tau_{\ell'})$ be the delay spread experienced by the signal, when reflected by the IRS. If the power
spectral density of the received signal  has an even symmetry around frequency $f_0$
and its bandwidth satisfies
\begin{equation} \label{eq:f_cond}
  B_w \leq \frac1{2 T_{\rm DS}}
\end{equation}
then the solution to~\eqref{eq:UCQP_no_mpath} will be given by~\eqref{eq:NB}.
\end{proposition}

  The proof is provided in~\ref{app:prop:3}.

In a 2D scenario where the antenna arrays and the IRS have a linear
shape and the IRS size is small compared to the BS--IRS and IRS--UE
distances, a special case of~\eqref{eq:hyp1} can be considered. Indeed,
under such hypothesis the dependence on $\ell$ of $K_{\ell}$ can be
dropped and the angles $\phi^{(1)}_{i,\ell}$ and
  $\phi^{(2)}_{j,\ell}$ (the directions of the $i$-th
  BS antenna and of the $j$-th UE antenna as observed from the $\ell$-th IRS element) are independent of the
  array indices, i.e. $\phi^{(1)}_{i,\ell}=\phi_{\rm BS}$ and
  $\phi^{(2)}_{j,\ell}=\phi_{\rm UE}$. Also, $\tau_{\ell}\approx C+\ell \tau$ where $C$ is a constant, 
  $\tau = \frac{\Delta (\sin \phi_{\rm BS} - \sin \phi_{\rm UE})}{c}$,
$\Delta$ being the IRS element spacing. Then, $T_{\rm DS}=\max_{\ell, \ell' =
  1,\dots,L} |\ell\tau - \ell'\tau| = (L-1)|\tau|$ and
condition~\eqref{eq:f_cond} becomes
\begin{equation}
     \label{eq:f_cond_1} B_w \leq \frac{c}{2(L-1)\Delta |\sin \phi_{\rm BS} - \sin \phi_{\rm UE}|}\,.  
\end{equation}
For example, for $\Delta=c/(2f_0)$ (i.e., the IRS element spacing
is half the wavelength $\lambda_0 = c/f_0$) and in the most unfavorable (albeit unrealistic)
geometric deployment of the network nodes (i.e., $\phi_{\rm BS}=\phi_{\rm UE}=\pi/2$), we have
  \begin{equation}
        \label{eq:f_cond_2} \frac{B_w}{f_0} \leq \frac1{2(L-1)}\,.
  \end{equation}
Note that the above expression is important since it directly relates
the normalized bandwidth $B_w/f_0$ to the inverse of the IRS size,
$L$.
 We also observe that the solution
  $\gammav^{\rm NB}(f_0)$ is easy to compute since it only requires
  some information about the geometry of the system, i.e., the delays
  $\tau_{\ell}$, and the value of the central frequency of the signal
  spectrum.

\subsubsection{Small BS and UE array size and discrete signal spectrum\label{sec:discrete_spectrum}}
If the transmitted signal has a spectrum organized in $S$  
subcarriers located at frequencies $f_s$, $s=1,\ldots, S$, the power spectral
density matrix $\Gm(f)$ can be written as
\[ \Gm(f) = \sum_{s=1}^S \Gm_s\delta(f-f_s) \]
where $f_s$ is the frequency associated to the $s$-th subcarrier.
Under the hypothesis made in Section~\ref{sec:small_arrays} (i.e., small BS, UE and IRS sizes and a 2D scenario)
the entries of matrix $\Tm$ in~\eqref{eq:UCQP_no_mpath} read as follows:
\begin{equation} \label{eq:case2}
[\Tm]_{\ell,\ell'} = K^2\sum_{s=1}^S M_s\ee^{\jj 2\pi (\ell'-\ell) \tau f_s} 
\end{equation} 
where $M_s = \trace\left\{\Am(f_s)\Gm_s\Am\Herm(f_s)\right\}$ is
proportional to the received signal power for subcarrier $s$. Hence
$\Tm = \sum_{s=1}^S M_s \zv_s\zv_s\Herm$,
$\zv_s=[z_{s,1},\ldots,z_{s,L}]\Tran$, and $z_{s,\ell}=K\ee^{-\jj 2\pi
  \ell \tau f_s}$.  In the case of a large number of IRS elements, we
state the following proposition:
\begin{proposition} \label{prop:4}
When $L \rightarrow \infty$, for a matrix $\Tm$ given by
\eqref{eq:case2}, the solution of the UCQP in \eqref{eq:UCQP_no_mpath} is given
by $\gammav^{\rm NB}(f_{s^*})$ where $s^*$ is the subcarrier index
maximizing over $s$ the term $M_s$ defined after~\eqref{eq:case2}, i.e., $s^* = \arg\max_s M_s$.
\end{proposition}

The proof is provided in~\ref{app:prop:4}. The evaluation of $\gammav^{\rm NB}(f_{s^*})$ requires to compute the
 terms $M_s$, which in turn requires the knowledge of the matrix
 $\Am(f)$, defined after \eqref{eq:hyp1}, and of the power spectral
 density $\Gm_s$.

 \subsubsection{Optimal narrowband solution\label{sec:NB_opt}}
 Finally, we consider the narrowband solution,
 $\gammav^{\rm NB}(f^*)$, obtained by searching in the band $\Bc$ 
 for the  frequency $f^*$ maximizing the received power, i.e.,
 \begin{equation}
   f^* = \arg \max_{f \in \Bc} (\gammav^{\rm NB}(f)) \Herm \Tm \gammav^{\rm NB}(f)\,. \label{eq:NB_opt}
 \end{equation}
 The optimal frequency $f^*$ can be obtained by plain numerical search.
 With respect to previous
 NB solutions, it requires an exhaustive search on the
 frequency space and the knowledge of the matrix $\Tm$ defined
 in~\eqref{eq:Tm}.

\section{Performance assessment\label{sec:results}}
 \begin{figure}[t]
  \centering
\includegraphics[width=1.0\columnwidth]{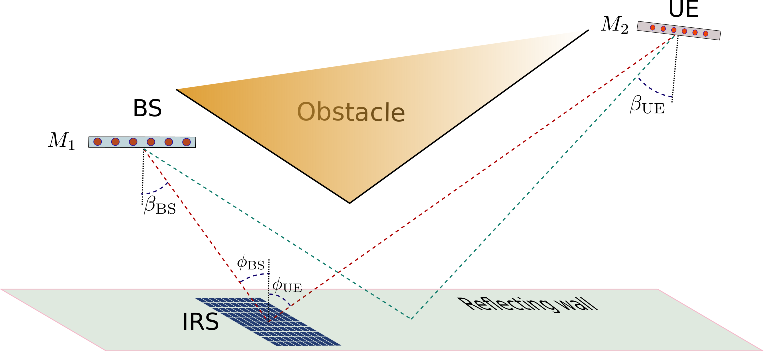}
\caption{Geometry of the simulated wireless communication
  network. The IRS is attached to a wall which also acts as a
  reflector. The IRS has square shape and is composed of $L=L_s^2$
  elements.}
\label{fig:network_simulation}
\end{figure}
 \begin{figure}[t]
  \centering
\includegraphics[width=1.0\columnwidth]{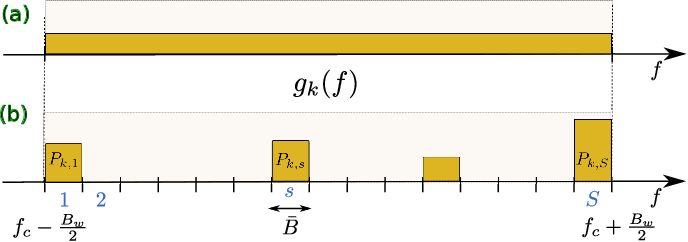}
\caption{Power spectral density profile of the
  signal transmitted by the $k$-th beam: (a) the entire available bandwidth is dedicated to a
  single user and (b) only some sub-bands are assigned to the UE.}
\label{fig:Gf}
 \end{figure}
 We now assess through simulation the performance of the IRS
 configuration options outlined in Section~\ref{sec:max_Pr}.  We
 specifically measure and discuss the influence that the system parameters
 (such as the IRS size, the presence of multipath components, and the
 signal bandwidth $B_w$) have on the achievable rate.
The next two subsections will be devoted to the description of the
simulation setup, i.e., the network geometric model, the channel characteristics
and the properties of the transmitted signal.

\subsection{Geometric model and channel characteristics\label{sec:results_model}}
In the simulations we consider the 2D scenario depicted in
Figure~\ref{fig:network} and a wireless communication network
operating in the THz band at central frequency $f_c$ = 300\,GHz.  The
BS is equipped with a uniform linear array (ULA) of $M_1$ antennas
spaced by $\Delta=\lambda_c/2$ where
$\lambda_c=c/f_c=1$\,mm is the wavelength corresponding to the central
frequency and $c$ is the speed of light. The UE has an ULA composed of
$M_2=8$ elements, spaced by $\Delta$.  Furthermore, for the BS we
assume 3GPP sectored antenna element~\cite{3gppchanmodel}
characterized by 8\,dBi maximum directional gain, whereas for the UE we
consider isotropic (i.e. 0dBi) antennas.  The IRS, located on a wall
which may also act as a reflector (see
Figure~\ref{fig:network_simulation}), has a square shape and is
composed of $L=L_s^2$ elements, with $L_s$  elements per side, spaced by $\Delta$.  The area of an element is, thus,
$A=\Delta^2$ (there are no gaps among elements). As for the
  frequency response of the IRS element, we employ the model
  in~\eqref{eq:zeta_ell_approx}, assuming that the IRS operational
  bandwith is greater than the signal bandwidth.  Since at present no
  practical models exist for IRS operating at $f_c$ = 300\,GHz, we
  consider a transfer function similar to that depicted
  in~\cite[Figs. 3(a) and 4(a)]{Behrooz2023}, where
  $\widetilde{\zeta}(f) \approx \rho(f)\ee^{\jj \alpha(f-f_c)}$, with 
  $\rho(f)\approx -1$\,dB its magnitude and $\alpha\approx -25^\circ/$GHz a real constant.

The BS and UE are located at $d^{(1)}$ = 8\,m and
$d^{(2)}$ = 15\,m, respectively,  from the the IRS center.  The attenuation
  due to atmospheric molecular absorption in the 230--370\,GHz band
  ranges from 1\,dB/km to tens of
  dB/km~\cite{ITU-Attenuation}. However, in such band and for the
  short distances involved in indoor applications, the molecular
  absorption loss is not a major issue and can be neglected. As for
the reflection properties of the wall, we adopt the complex reflection
coefficient characterization of plasterboard panels~\cite{reflection}
depicted in~\cite[Figure 2]{noiTWC}.  Finally, at the UE, the noise power
spectral density is set to $N_0=-174$\,dBm/Hz.

\subsection{Power spectral density of the transmitted signal and beamforming vector\label{sec:results_G}} We assume that the
  BS transmits a signal of bandwidth $B_w$ whose support is
  $\Bc \subseteq [f_c-\frac{B_w}{2}, f_c+\frac{B_w}{2}]$.
  For every frequency $f\in \Bc$, the BS employs the antenna array to
  generate up to $K$ beams so that the transmitted signal can be
  described by the power spectral density matrix
\begin{equation}
  \Gm(f) = \sum_{k=1}^K g_k(f) \vv_k(f)\vv_k\Herm(f)\label{eq:G_simulations}
\end{equation}
where $g_k(f)$ is the scalar power spectral density associated to the
$k$-th beam, whose support is
$\Bc_k\subseteq \Bc$ for
$k=1,\ldots,K$ (see the examples depicted in Fig.~\ref{fig:Gf}).  So, the rank of $\Gm(f)$ can be up to $K$,
provided that the dimension of the vector space generated by the set
of beamformers, $\vv_k(f)$, is $K$.  Then, according to~\eqref{Eq:Pt},
the total transmit power is given by
$P^{\rm tx} =\sum_{k=1}^K P^{\rm tx}_k$, where$P^{\rm tx}_k=\int_{\Bc_k} g_k(f)\dd f$ is the
transmit power associated to the $k$-th beam. So
far, there are no standards regulating communications at such high
frequencies, however the formula in~\eqref{eq:G_simulations} is
general enough to encompass many scenarios that can be envisioned, as
those described in~\cite{Shafie2019}. For example, and referring to the $k$-th beam:
\begin{itemize}
\item the entire available bandwidth is dedicated to the UE as represented in Fig~\ref{fig:Gf}(a). This is the case
  of single-carrier UWB signaling, obtained e.g. by
  transmitting very narrow pulses. If the spectrum is flat, the scalar power
  spectral density $g_k(f)$ can be expressed as
  $g_k(f)=P^{\rm tx}_k/B_w$;
\item only portions of the available bandwidth are dedicated
  to the UE, as depicted in Fig~\ref{fig:Gf}(b), where the spectrum is
  divided in $S$ sub-bands, each having the same bandwidth $\bar{B}$,
  i.e, $S \bar{B}= B_w$. One or more (not necessarily adjacent)
  sub-bands are allocated to the UE. In a scenario where
    multiple UEs compete for spectrum resources, this choice depends on
    the data rate required by the UE and on the allocation strategy of
    the sub-bands.

    In the example of Figure~\ref{fig:Gf}(b), four
    sub-bands are assigned to the UE through the power spectral density $g_k(f)$. The area of the yellow
    rectangles represents the power associated to each sub-band.
    In this scenario the scalar power spectral density $g_k(f)$ can be written as
    \begin{equation}
      g_k(f) = \sum_{s=1}^S \frac{P^{\rm tx}_{k,s}}{\bar{B}} \Pi_{\bar{B}}(f-f_s)\label{eq:gk}
    \end{equation}
    where $\Pi_{\bar{B}}(f)=1$ for $|f|\le \frac{\bar{B}}{2}$ and 0
    elsewhere, and $f_s=f_c+\bar{B}(s-(S+1)/2)$, $s=1,\ldots,S$ is the
    central frequency of the $s$-th sub-band.  Finally, $P^{\rm tx}_{k,s}$ is
    the power associated by the transmitter to beam $k$ in sub-band $s$. Note that $P^{\rm tx}_{k,s}=0$
    if the sub-band $s$ is not allocated to the UE for beam $k$. The power of the $k$-th beam is given by $P^{\rm tx}_k=\sum_{s=1}^S P^{\rm tx}_{k,s}$. 

    Let $\Sc_k$ be the set of sub-bands assigned to the UE for beam $k$, having cardinality $N_k=|\Sc_k|$. In the
    simulations we consider two possible power allocation scenarios:
    \begin{itemize}
      \item all allocated sub-bands have the same
        power, i.e. $P^{\rm tx}_{k,s}=P^{\rm tx}_k/N_k$ for $s\in \Sc_k$ and $P^{\rm tx}_{k,s}=0$
        elsewhere; we will refer to this allocation as ``equal power loading'';
      \item the powers $P^{\rm tx}_{k,s}$, $s\in \Sc_k$ are
        obtained by extracting outcomes of independent uniformly distributed random variables 
       normalized so that their sum is $P^{\rm tx}_k$. We will refer to this allocation as ``random power loading''.     
    \end{itemize}  
\end{itemize}

The transmitter array can also generate beam squint. However, its
effects can be mitigated, e.g.,  through frequency-dependent
architectures, which adopt virtual arrays and TTD
lines~\cite{WB_beamforming_mMIMO}, so as to achieve performance very
close to fully digital transceivers. Therefore, for the beamforming
vector, we can consider the two following techniques at the BS:
\begin{itemize}
\item a single frequency-independent beamforming vector tuned on a specific
  frequency, say the central frequency $f_c$, whose $m$-th entry, $m = 0,\dots,M_1-1$, is given by
  \begin{equation}
    [\vv_k(f)]_m = \frac{1}{\sqrt{M_1}}\ee^{\jj 2 \pi m (\Delta/c) f_c \sin(\beta_{{\rm BS},k})}\label{eq:central_bf}
  \end{equation}
  for  $f\in \Bc$, where $\beta_{{\rm BS},k}$ is the direction of beam $k$ at central
  frequency $f_c$, as observed from the center of the BS array.  When
  many beams are generated, the angles $\beta_{{\rm BS},k}$ should
  significantly differ, in order to grant spatial diversity and a rank
  $K>1$.  The beamformer in~\eqref{eq:central_bf} is simple to
  implement since it is independent of $f$. However, when wideband
  signals are transmitted, it generates beam squint, which makes the
  beam direction frequency-dependent. In the following, we will refer
  to~\eqref{eq:central_bf} as ``Central beamforming'' technique;
\item a set of $S$ beamforming vectors tuned on the central frequencies $f_s$ of the sub-bands, $s{=}1{,}{\ldots}{,}S$, whose $m$-th entry, $m{=}0{,}\dots{,}M_1{-}1$, is given by
  \begin{equation}
    [\vv_k(f)]_m = \frac{1}{\sqrt{M_1}}\ee^{\jj 2 \pi m (\Delta/c) f_s \sin(\beta_{{\rm BS},k})}\,,\label{eq:adapted_bf}
  \end{equation}
  for $f\in [f_s{-}\bar{B}/2, f_s{+}\bar{B}/2]$
  and $s{=}1{,}\ldots{,}S$. This beamforming technique will be referred to as ``Adapted
  beamforming''. 
  It can be implemented with, e.g., a fully digital transmit array, and can significantly reduce the beam squint effect
  generated by the BS array (provided that $\bar{B}$ is sufficiently
  small), at the expense of a much higher transmitter complexity. 
\end{itemize}

\begin{figure}[t]
  \centering
  \includegraphics[width=0.8\columnwidth]{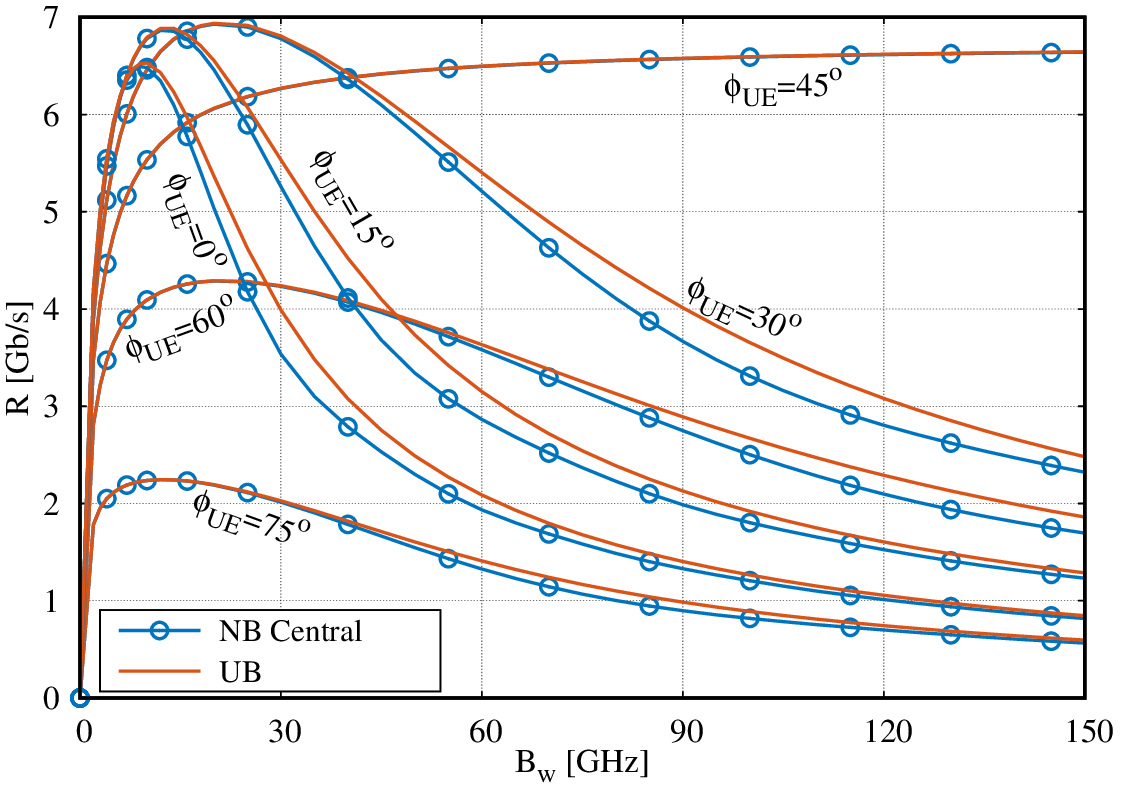}
\caption{Rate, $R$, versus $B_w$, as $\phi_{\rm UE}$ varies, for $K=1$, $M_1=16$,
  $P^{\rm tx}=22$\,dBm, $L=64 \times 64$ and $\phi_{\rm BS}=45^\circ$.  The
  BS implements single carrier UWB communication with
  ``Adapted beamforming''.}
\label{fig:spread1}
 \end{figure}

\subsection{IRS configuration solutions and complexities}
We assess the performance
of the IRS configuration options described in the previous sections, in particular:
\begin{itemize} 
\item the narrowband (NB) solution tailored to the central frequency
  $f_c$, $\gammav^{\rm NB}(f_c)$, whose elements are as
  in~\eqref{eq:NB} with $f_0=f_c$ and $\tau_\ell=d_\ell/c$,
  according to the geometry depicted in
  Figure~\ref{fig:network_simulation}. This solution  is the simplest one since it has the advantage
  of depending only on the system geometry, i.e., on the position of
  the BS and UE, while being independent of the signal spectrum. Such solution was
    also proposed in~\cite{BeamSquintMitigating} and reported
    to provide good performance for bandwidths up to 7\%. However, this is not the case for much larger bandwiths,
    as we will show in Section~\ref{eq:results_rank1}.  In
  the following it will be referred to as ``NB Central''.
  The complexity of computing $\gammav^{\rm NB}(f_c)$ is about $O(L)$; 
\item for spectra as in Fig.~\ref{fig:Gf}(b), the narrowband
  solution $\gammav^{\mathrm{NB}}(f_{s^*})$ defined in
  Proposition~\ref{prop:4};
  this solution requires to evaluate the received power for each sub-band, and the knowledge of
  the channel matrix. In the following it is referred to as ``NB
  Max-power''. Its complexity is about $O(\max_kN_kL)$
  where $N_k$ is the number of allocated sub-bands for beam $k$;
\item the narrowband solution
    $\gammav^{\rm NB}(f^*)$, described in Section~\ref{sec:NB_opt} and
    in the following labeled ``NB Optimum''. We recall that such
    technique requires to search in the frequency space $\Bc$. To
    reduce complexity, in the simulations, we discretized the
    frequency space and tested only the central frequency of each
    sub-band of width $\bar{B}$. Therefore, the complexity of computing
    $\gammav^{\rm NB}(f^*)$ is about $O(SL^2)$ where $S$ is the number
    of sub-bands contained in the frequency range $[f_c-\frac{B_w}{2},f_c+\frac{B_w}{2}]$, and $S = B_w/\bar{B}$;
  \item the outcome of the ``SCF'' algorithm proposed in~\cite{Aldayel17},
    which numerically solves the UCQP problem~\eqref{eq:UCQP}. It works iteratively
    and requires the knowledge of the matrix $\Tm$ and of the vector
    $\qv$. Its complexity is $O(L^3 + FL^{2.373})$ where
    $F$ is the number of iterations (sometimes large) required to reach convergence. In general,
    the convergence rate depends on the value of the threshold used to
    declare convergence and on the problem size, $L$.  Also, the
    outcome of SCF depends on the starting point of the iterative
    procedure, i.e., an initial random guess of the vector
    $\gammav$. However, due to the non-convexity of UCQP, not all
    starting points lead to the same solution. Then, we decided to
    run the algorithm several times for each instance of
    the problem~\eqref{eq:UCQP} and keep the best solution found;
  \item {\color{black}the solution provided by semidefinite
    relaxation (SDR) techniques described
    in~\cite{SemidefiniteRelaxation} and also proposed
    in~\cite{Wu-Zhang2018} to optimize IRS-aided communications in the presence of narrowband signals, independently of the transmitted signal spectrum;}
 \item the vector $\gammav^\star$ obtained by extracting the phases of
   the eigenvector $\uv^{\max}$ corresponding to the maximum
   eigenvalue of the matrix $\Tm$, as specified
   in~\eqref{eq:eig}. This technique requires the knowledge of the
   matrix $\Tm$ and will be referred to as ``Max-eig phase''; its
   complexity
   is about $O(L^3)$;
\item the upper bound in~\eqref{eq:upper_bound3},
  which will be used as a benchmark and denoted by the label ``UB''.
\end{itemize}

\subsection{Numerical results for a rank-1 matrix $\Gm(f)$\label{eq:results_rank1}}
We start by considering a matrix $\Gm(f)$ having rank-1 on its support
($K=1$) and a single-carrier UWB signal while neglecting both
shadowing and multipath. In this case, the power spectral density of
the transmitted signal can be represented as in Fig~\ref{fig:Gf}(a).
Figures~\ref{fig:spread1}
reports the achievable rate $R$ (Gb/s), computed by using~\eqref{eq:R},
plotted versus signal bandwidth, $B_w$, for $M_1=16$ transmit
antennas, transmit power $P^{\rm tx}=22$\,dBm,
$\phi_{\rm BS}=45^\circ$, and $L_s=64$ (hence $L=L_s^2=4096$ IRS
elements).  In Figure~\ref{fig:spread1} the BS employs the
``Adapted beamforming'' technique so as to eliminate the beam-squint
effect at the BS. The figure reports 6 sets of curves, characterized
by the angles
$\phi_{\rm UE}\in \{0^\circ,
15^\circ,30^\circ,45^\circ,60^\circ,75^\circ\}$.  Each set is composed
of two curves: the rate achieved by using ``NB central'' technique,
and the upper bound ``UB''. First of all, we observe that ``NB
Central'', despite its simplicity, shows excellent performance, very
close to the upper bound ``UB''.  We also note that for
$\phi_{\rm BS}=\phi_{\rm UE}=45^\circ$ both incident and reflected
angles are the same and hence the IRS acts as an ordinary mirror if we
impose a constant phase-shift, $\theta_\ell$, across the surface. In
this particular situation the IRS does not generate any beam-squint
effect and the rate increases with $B_w$ according to the well-known
law $B_w\log_2(1+P^{\rm rx}/(N_0B_w))$.  However, when the incident
and reflected angles differ, the direction of the beam generated by
the BS is frequency-dependent and, hence, part of the signal energy is
spread in unwanted directions. This causes a rate loss, which increases
as $B_w$ increases.  For any $\phi_{\rm UE} \neq \phi_{\rm BS}$, the
rate curve shows a peak whose maximum is higher for
$\phi_{\rm UE}<45^\circ$. This is due to the fact that the effective
area of an IRS element, $A^{(2)}_{j,\ell}$ depends on
$\cos(\phi^{(2)}_{j,\ell})$ and, hence, is maximized when the UE is
approximately located in front of the IRS. In the scenario depicted in
the figure, communication does not incur severe performance losses
if the bandwidth is about 5\% to 15\% of the central frequency,
although some tolerance is allowed depending on the system
geometry. The ``UB'' curves tell us that, when the signal bandwidth,
$B_w$, exceeds 30\,GHz, penalties due to IRS beam squint severely
reduce the achievable rate, no matter how the IRS is configured.
Note that for each angle $\phi_{\rm UE}$, there is an optimal
  $B_w$ that maximizes the rate. This means that, in order to avoid rate
  penalties, the signal bandwidth must be swiftly adapted to the system
  geometry especially in high mobility scenarios.

\begin{figure}[t]
    \centering 
  \includegraphics[width=0.8\columnwidth]{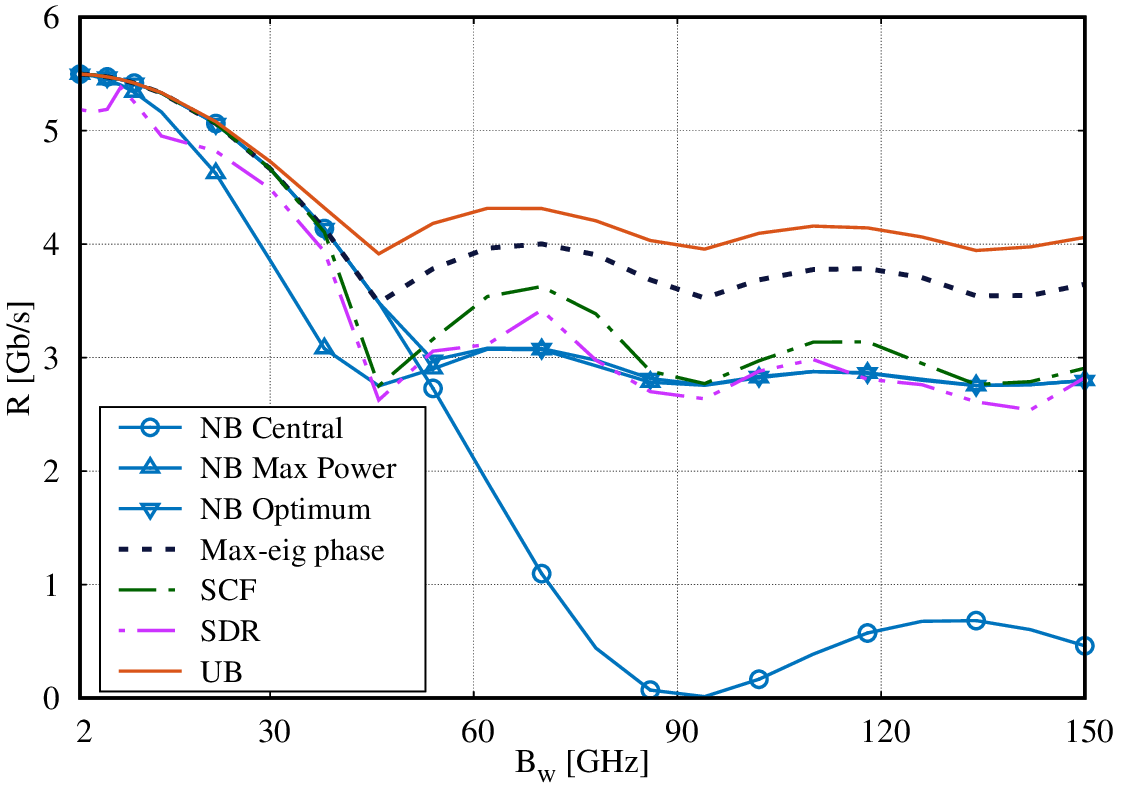}
\caption{Rate $R$ versus $B_w$, for $K=1$, $M_1=16$, $L=64\times 64$, $\phi_{\rm BS}=45^\circ$,
  $\phi_{\rm UE}=30^\circ$, $P^{tx}_{1,s}=22$\,dBm,
  ``equal power loading'' among sub-bands, and ``Adapted beamforming''
  at the BS. The signal power spectral density is
  organized as in Figure~\ref{fig:Gf}(b) with $N_1=2$.
  The effects of multipath and
  shadowing are not taken into account.}
\label{fig:sub_bands1}
\end{figure}

 In Figure~\ref{fig:sub_bands1} we show the achievable
  rate plotted versus the available bandwidth in the case where the
  spectrum is organized in $S$ sub-bands, each having
  bandwidth $\bar{B}=1$\,GHz, out of which $N_1=4$ have been allocated
  to the UE
  so that the support of the signal spectrum has measure
  2\,GHz.
  The sub-bands are organized as depicted in Figure~\ref{fig:Gf}(b):
  the innermost ones are randomly located
  in the available slots. The curves have been obtained
  by averaging over 100 realizations of such allocations.
  The transmitted power per allocated
  sub-band is set to  $P^{\rm tx}_{1,s}=22$\,dBm
  and equal power loading across sub-bands is
  employed. The BS and the UE are observed from the IRS under the
  angles $\phi_{\rm BS}=45^\circ$ and $\phi_{\rm UE}=30^\circ$ (see
  Fig.~\ref{fig:network_simulation}). Again,  shadowing and
  channel multipath components are neglected.
\begin{figure}[t]
   \centering
  \includegraphics[width=0.8\columnwidth]{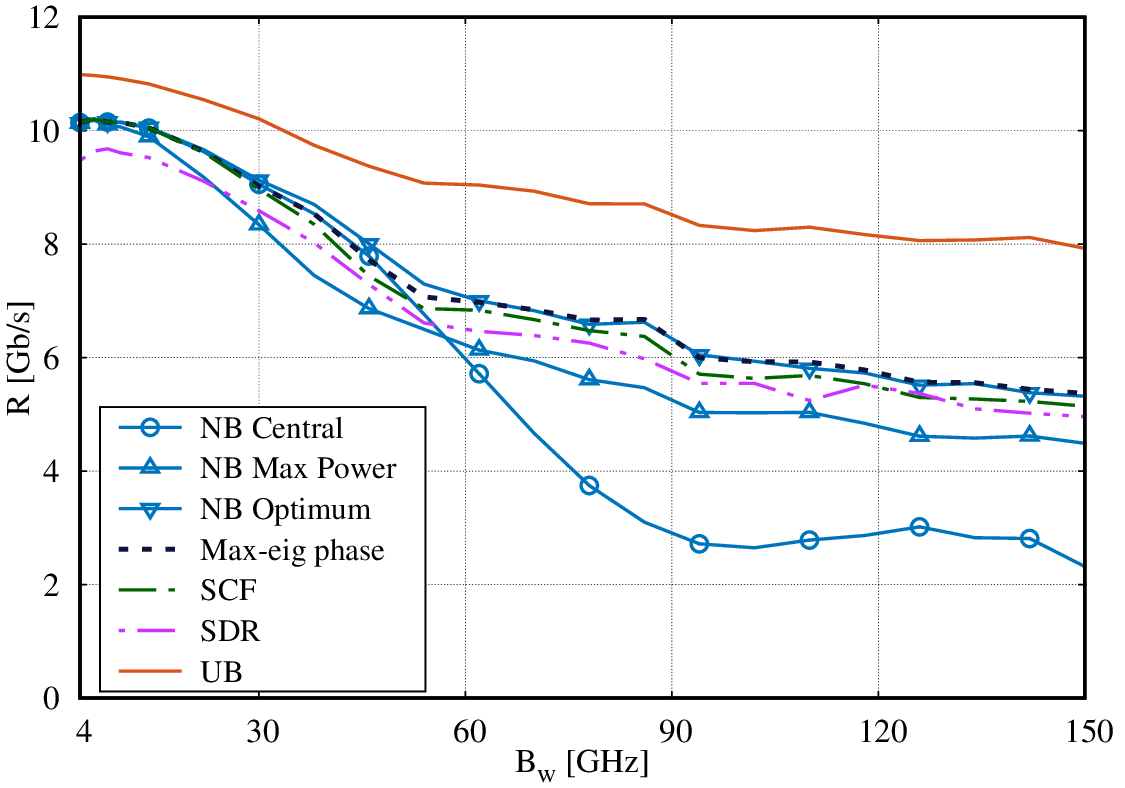}
  \caption{Rate $R$ versus $B_w$, for $K=1$, $M_1=16$,
    $L=64\times 64$, $\phi_{\rm BS}=45^\circ$,
    $\phi_{\rm UE}=30^\circ$, $P^{\rm tx}_{1,s}=22$\,dBm per sub-band,
    ``random power loading'' among sub-bands and ``Adapted
    beamforming'' at the BS.  The signal power spectral density is
    organized as in Figure~\ref{fig:Gf}(b) with
    $N_1=4$. 
    The effects of multipath and shadowing are neglected.}
\label{fig:sub_bands1_random_loading}
\end{figure}
 As can be observed, the ``NB central'' solution performs very close to the
upper bound for $B_w\le30$\,GHz, whereas it suffers significant
performance losses for $B_w\ge 45$\,GHz and  even drops to zero for
$B_w=90$\,GHz.  This oscillating effect is due to the radiation
pattern of the IRS. A rate close to zero occurs when the direction of
the UE, as observed from the IRS, corresponds to a null between two
adjacent side lobes.  Interestingly, the other IRS configuration
techniques do not seem to suffer from this effect although some small
oscillations can still be observed in the rate.  The ``NB Max power''
and ``NB Optimum'' techniques, although suboptimal for $B_w<30$\,GHz,
provide almost constant rate for larger $B_w$.  The best
performance is provided by ``Max-eig phase'' which remains close to
``UB'' for all the considered bandwidths.
{\color{black} The solution obtained through SDR 
performs similar to ``NB Optimum'' and significantly worse than ``Max-eig phase''.}
 Finally, we observe that the SCF
algorithm does not provide excellent performance but this is not
surprising. Indeed, although it is designed for solving the problem
in~\eqref{eq:UCQP}, it is not guaranteed that its solution translates into an achievable rate higher than that provided by other
techniques. In fact, the maximizer of~\eqref{eq:UCQP} maximizes the
upper bound to the rate, not the actual rate.

\begin{remark}Please note that, for all considered techniques, the
  best performance is achieved when $B_w$ is as small as
  possible. This corresponds to the case where the $N_1$ allocated
  sub-bands are adjacent (i.e., $B_w=N_1\bar{B}$). However, we point
  out that due to system constraints, this situation might not occur
  in a scenario where many users compete for the same channel
  resources. In general, an evaluation of the trade-off between the
  benefit of having additional allocated sub-bands, and the drawback
  of larger $B_w$ needs to be carefully addressed at system level.
  \end{remark}

For the same parameters as in Fig.~\ref{fig:sub_bands1},
Fig.~\ref{fig:sub_bands1_random_loading} shows the achievable rate
when random power loading is employed across sub-bands, for $N_1=4$.
In this case ``Max-eig phase'' shows performance similar to ``NB optimum''
and outperfors all the other techniques whereas
``NB Central'' provides slightly better performance than
for $N_1=1$, although still being the worst technique for $B_W{>}60$\,GHz. 

\begin{figure}[t]
   \centering
  \includegraphics[width=0.8\columnwidth]{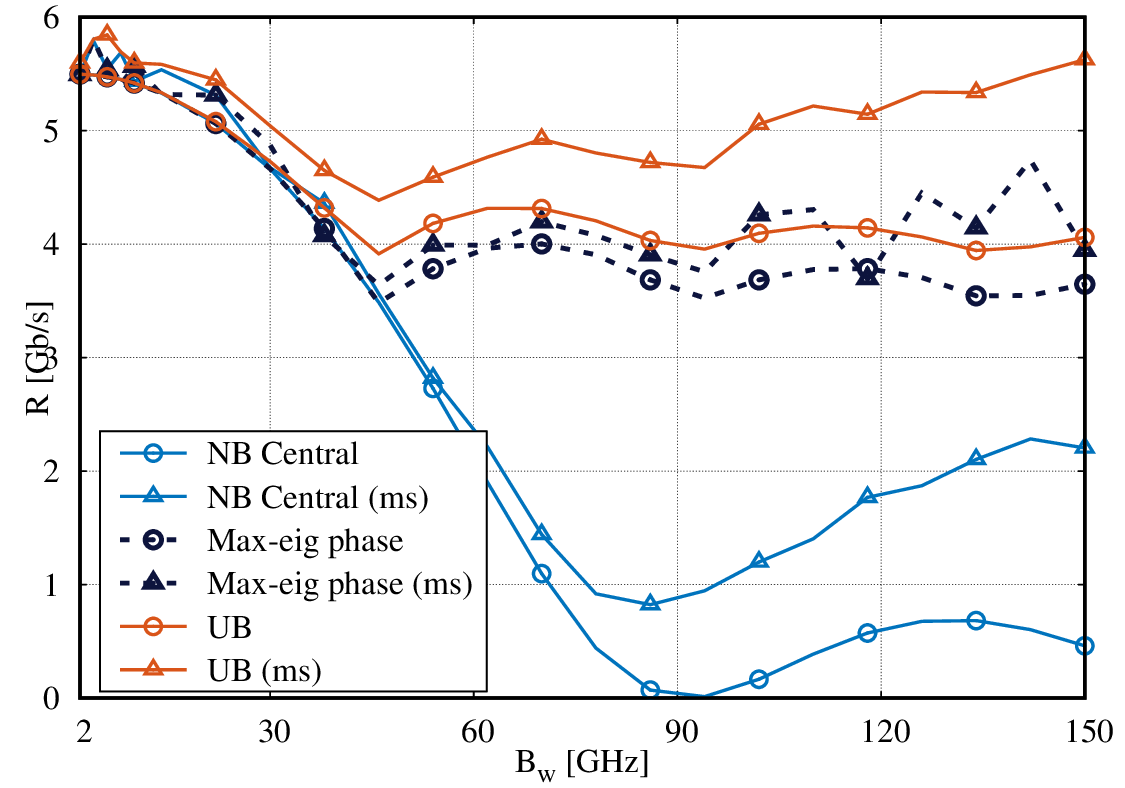}
\caption{Rate $R$ versus $B_w$, for $K=1$, $M_1=16$, $L=64\times 64$, $\phi_{\rm BS}=45^\circ$, $\phi_{\rm
    UE}=30^\circ$, $P^{\rm tx}_{1,s}=22$\,dBm per sub-band, ``equal
  power loading'' and ``Adapted beamforming'' at
  the BS.   The signal power spectral density is organized as
  in Figure~\ref{fig:Gf}(b) with $N_1=2$. 
  The effect of shadowing and multipath (denoted as ``ms'' in the legend) is considered
  (triangle markers) or neglected (circle markers).}
\label{fig:sub_bands_mixed}
\end{figure}

In Figure~\ref{fig:sub_bands_mixed} we highlight the effect of
multipath, i.e. the reflection by the wall, and shadowing (denoted as ``ms'' in the legend) on the rate performances,
for the
same system parameters as in Figure~\ref{fig:sub_bands1}. As for the
shadowing affecting the
BS--IRS-UE and BS--wall--UE paths, we considered independent realizations of a lognormal distribution with variance
$\sigma_{\rm sh}=2$.  As expected, the contribution of
the signal reflection by the wall is beneficial to the performance since it
carries additional energy to the receiver.  The rate increase is,
however, limited to about 5\%--10\% when ``Max-eig phase'' technique
is employed. Although ``NB central'' seems to be more benefitted by the signal reflection by the wall, with up to 70\,\% rate increase in some cases, it shows,
however, poor performance compared to ``Max-eig phase''.

\begin{figure}[t]
   \centering
  \includegraphics[width=0.8\columnwidth]{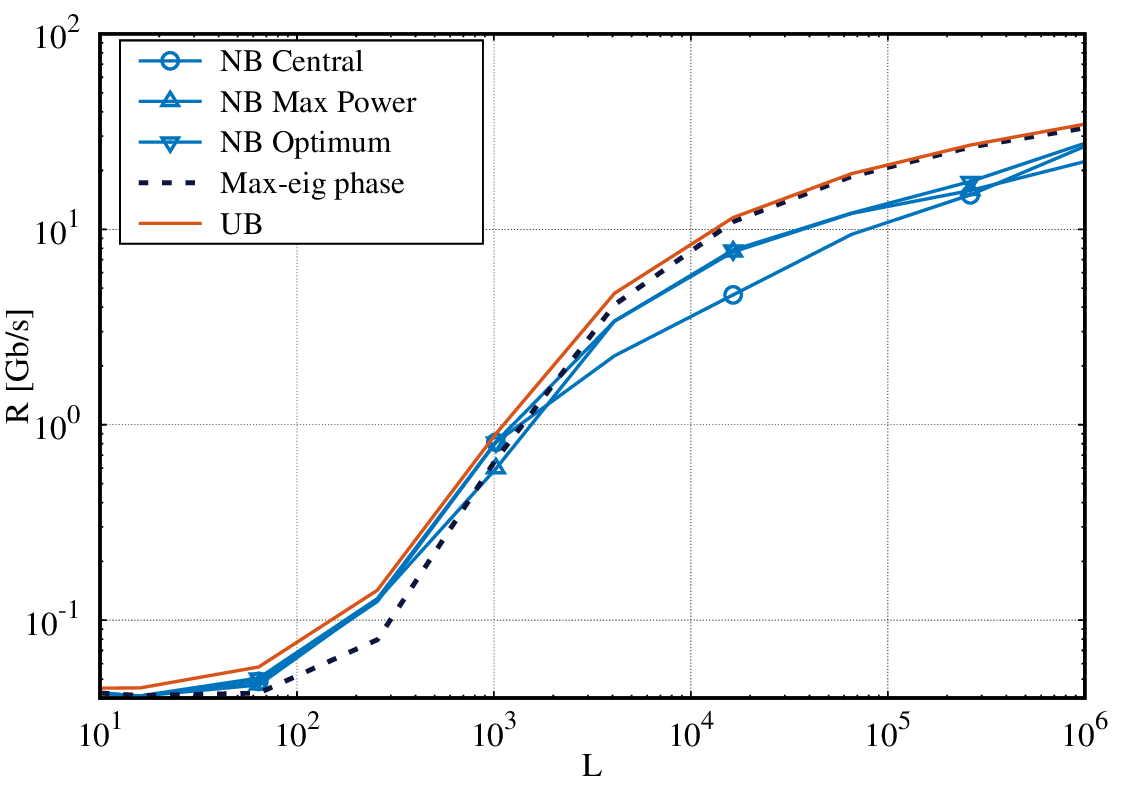}
\caption{Rate $R$ versus the IRS size, $L$, in the
  presence of multipath and shadowing, for $K=1$, $M_1=16$, $B_w=60$\,GHz, $\phi_{\rm
    BS}=45^\circ$m  $\phi_{\rm UE}=30^\circ$, $P^{\rm tx}_{1,s}=22$\,dBm per sub-band,
  ``equal power loading'' is adopted among the sub-bands and ``Adapted beamforming'' at the BS. The signal power
  spectral density is organized as in Figure~\ref{fig:Gf}(b) with
  $N_1=2$.
   The shadowing lognormal distribution has variance
  $\sigma_{\rm sh}=2$\,dB.}
\label{fig:sub_bands_L1}
\end{figure}

Figure~\ref{fig:sub_bands_L1} shows the rate $R$ versus the IRS size
$L$ in the presence of multipath and shadowing, modeled as in
Figure~\ref{fig:sub_bands_mixed}, for $M_1=16$, $B_w=60$\,GHz,
$\phi_{\rm BS}=45^\circ$ and $\phi_{\rm UE}=30^\circ$.
We observe that, 
for small IRS size, the ``NB'' techniques perform all very similarly
and their performances are close to those achieved by ``UB'' , whereas
they all tend to degrade for medium-to-large surfaces, i.e., $L>1000$.
Instead, the ``Max-eig phase'' technique performs better for large
surfaces and shows some degradation for $L<1000$.  The same behavior
can be observed in Figure~\ref{fig:sub_bands_L1}(right), where
$N_1=4$, although performance differences tend to reduce.

\begin{figure*}[t]
  \includegraphics[width=0.33\textwidth]{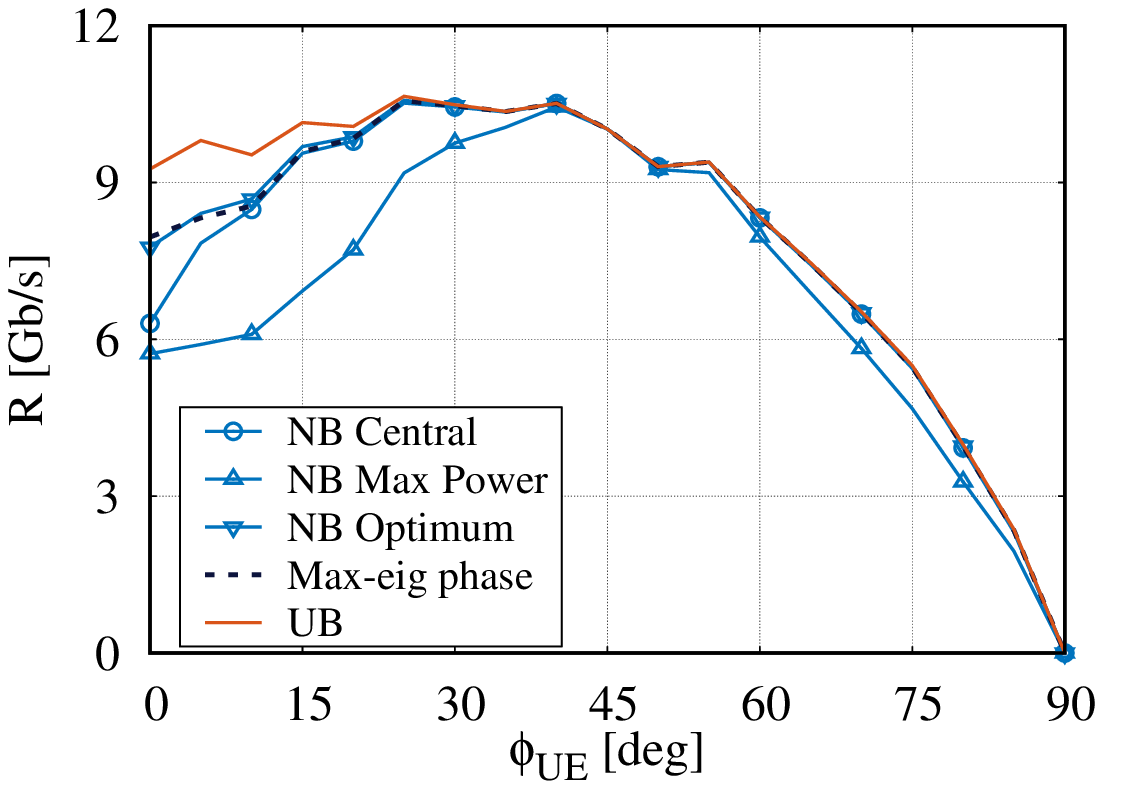}\includegraphics[width=0.33\textwidth]{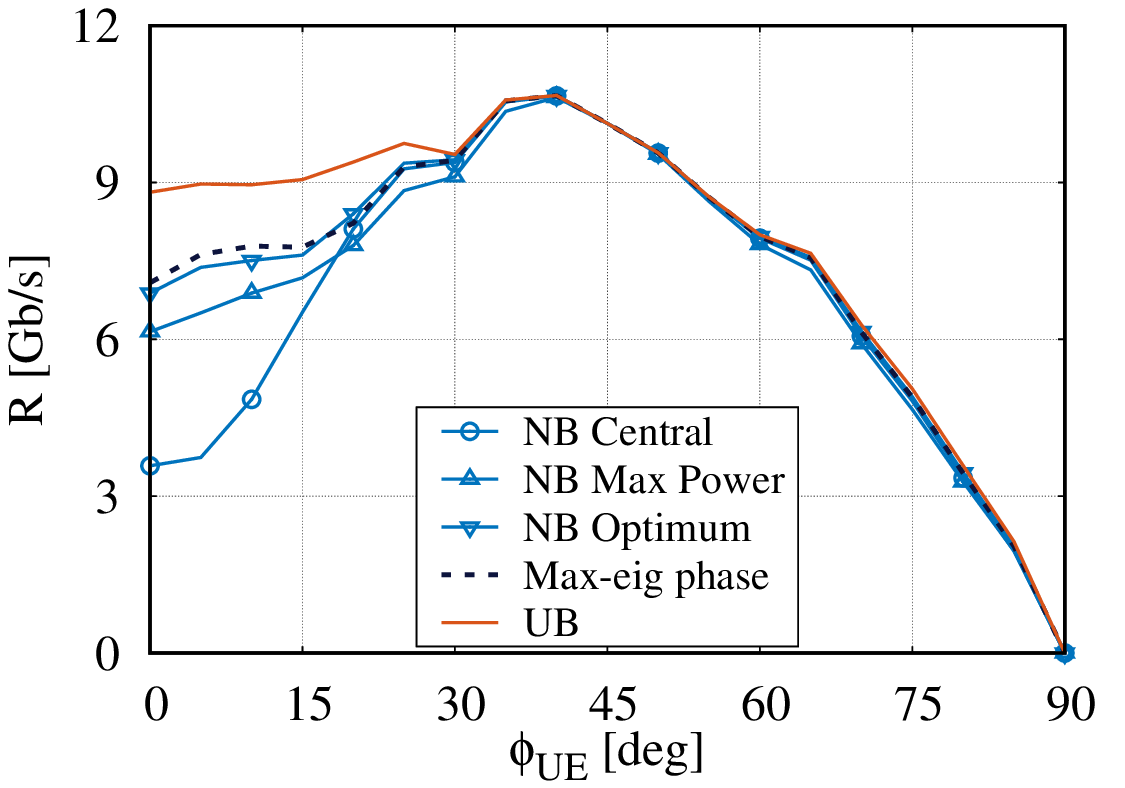}\includegraphics[width=0.33\textwidth]{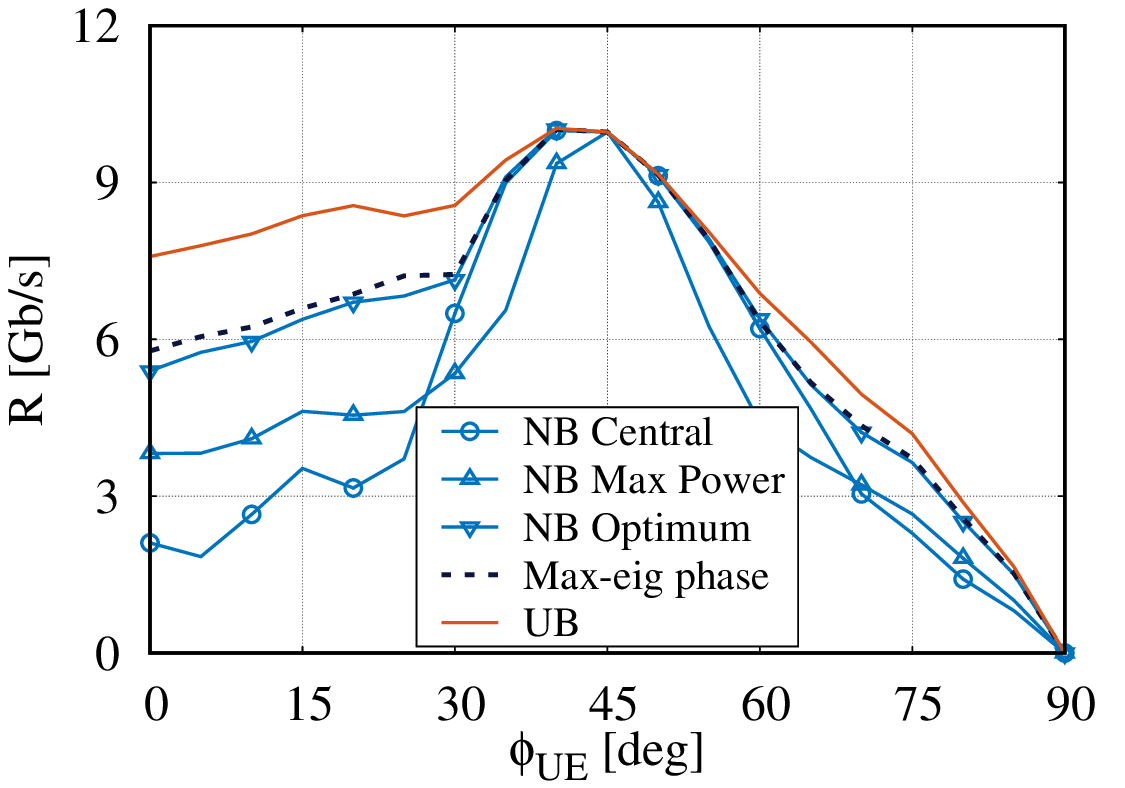}
\caption{Achievable rate, $R$, versus angle $\phi_{\rm UE}$ for
  $\phi_{\rm BS}=45^\circ$ and $B_w \in \{18,30,60\}$\,GHz from left
  to right. $N_1=4$ sub-bands are allocated to the single UE ($K=1$) with ``equal power loading'' criterium  and ``Adapted beamforming'' technique is adopted at the BS. Transmit power per sub-band is $P^{\rm tx}_{1,s}=22$\,dBm.  The shadowing lognormal distribution has variance
  $\sigma_{\rm sh}=2$\,dB and there is no multipath. }
\label{fig:sub_bands_angle}
\end{figure*}

In Fig.~\ref{fig:sub_bands_angle} we assess the influence of the
system geometry on performance by measuring the achievable rate as a
function of the angle $\phi_{\rm UE}$ (see
Fig.~\ref{fig:network_simulation}), for $M_1{=}16$, $N_1{=}4$ allocated
sub-bands, $P^{tx}_{1,s}$=22\,dBm in each sub-band, ``equal power
loading'' criterium on the sub-bands, shadowing with variance
$\sigma_{\rm sh}=2$\,dB and no wall reflection. The three figures
refer to the cases $B_w{=}20$\,GHz (left), $B_w{=}30$\,GHz (center) and
$B_w{=}60$\,GHz (right).  We first observe that, for
$\phi_{\rm UE}=45^\circ$,  all techniques behave the same, since the IRS
acts as an ordinary mirror. Also, as $\phi_{\rm UE}$ approaches
$90^\circ$ the rate tends to zero since the effective area of the IRS
towards UE, varying with the cosine of $\phi_{\rm UE}$, tends to
vanish. In general, in the range
$45^\circ \le \phi_{\rm UE}\le 90^\circ$ all techniques are almost
equivalent, whereas for smaller angles there are significant difference
among the performances they provide. The asymmetry of the curves
w.r.t. $\phi_{\rm UE} = 45^\circ$ is explained again by the dependency
of the IRS effective area towards the UE by the cosine of
$\phi_{\rm UE}$.
Also in this case, the ``Max-eig phase'' technique always outperforms
all the others, although ``NB Optimum'' provides the best trade-off
between performance and implementation complexity.

 \subsection{Numerical results for a rank-2 matrix $\Gm(f)$} 
\begin{figure*}[t]
\includegraphics[width=0.33\textwidth]{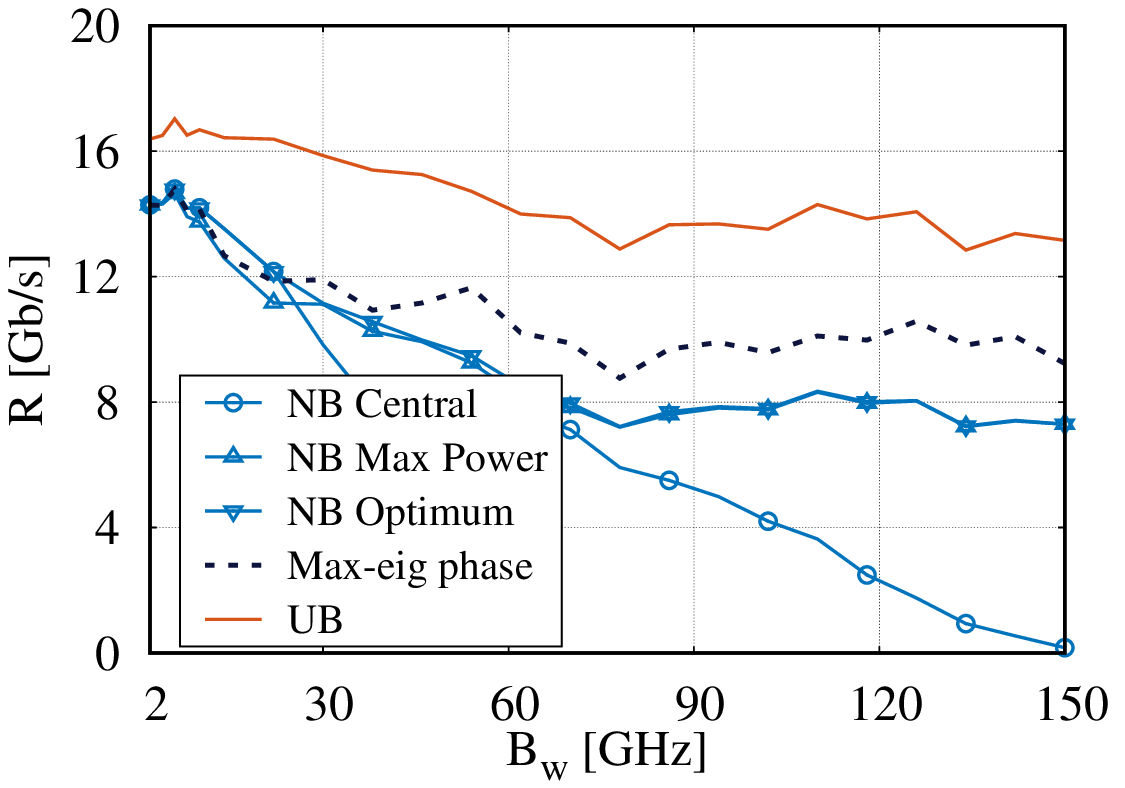}%
\includegraphics[width=0.33\textwidth]{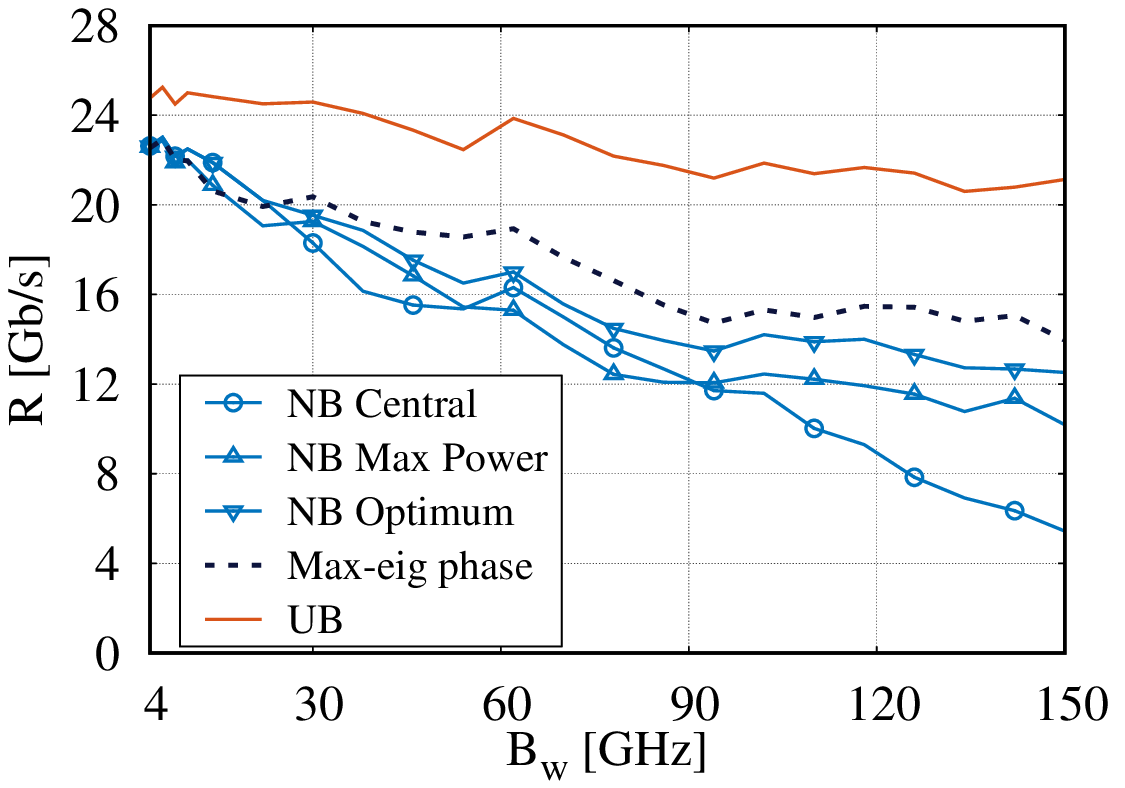}%
\includegraphics[width=0.33\textwidth]{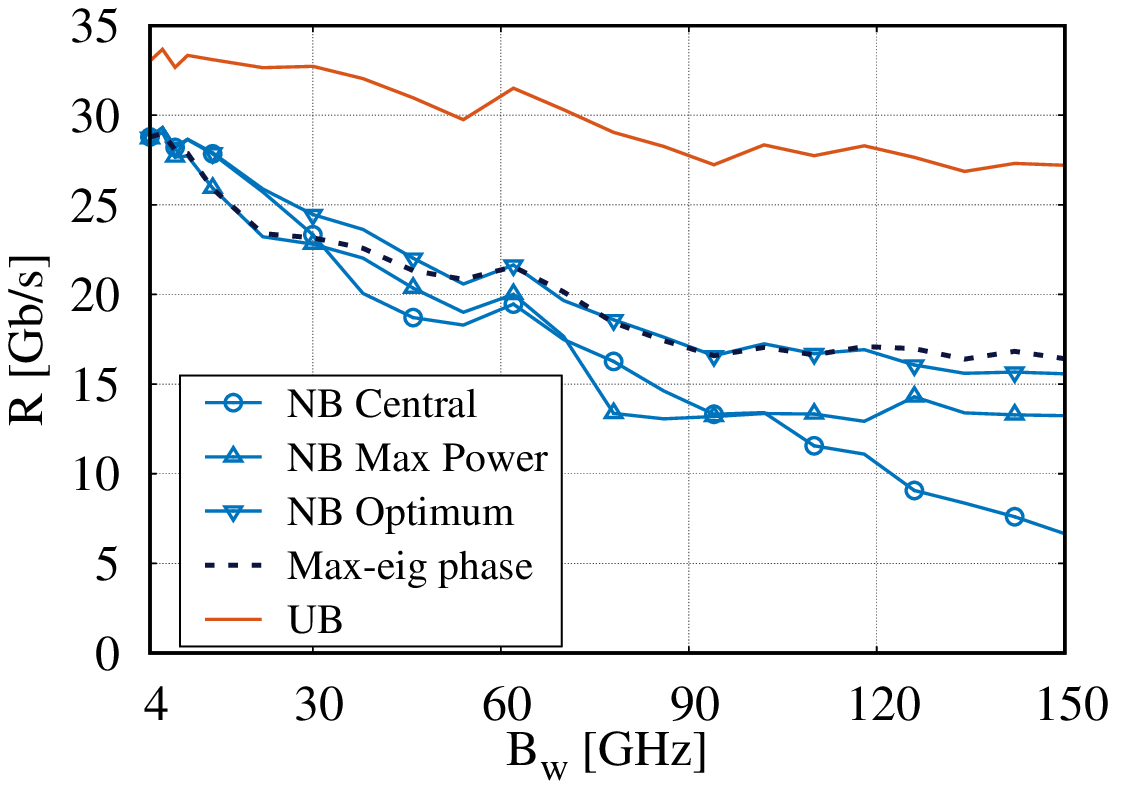}
\caption{Achievable rate, $R$, versus the  available bandwidth $B_w$ for a
  rank $K=2$ matrix $\Gm(f)$, $M_1=32$, transmit power 22\,dBm for each sub-band and each beam,
  with ``equal power loading'' among the sub-bands allocated to the user,  and ``Adapted
  beamforming'' at the BS. $N_1=N_2=2$ (left), $N_1=4$, $N_2=2$ (center) and
  $N_1=4$, $N_2=4$ (right).}
\label{fig:sub_bands_K2}
\end{figure*}

 Fig.~\ref{fig:sub_bands_K2} show the results of simulation assuming a power spectral density as in~\eqref{eq:G_simulations}, organized in
 sub-bands, with
 rank $K=2$. This means that, for any $f\in \Bc$, the BS generates up
 to $K=2$ beams.  In order to support a rank-2 power spectral density
 matrix, the channel is required to have enough spatial diversity to
 grant a rank-2 matrix $\Qm(f) = \Hm(f)\Gm(f)\Hm(f)\Herm$ as
 well. Then, the beamformers $\vv_k(f)$ defined
 in~\eqref{eq:G_simulations} should generate  sufficiently spatially separated beams for all
 $f\in \Bc$. For this reason, the results depicted in
 Fig.~\ref{fig:sub_bands_K2} refer to two IRSs of size $64
 \times 64$ elements, both located on the wall and separated by 5\,m from each
 other. These two IRSs, although physically separated, have
 to be considered as a single {\em logical} IRS of size $L =128\times
 64$ elements. Hence, the analytic derivations presented in this work
 are still valid.  In Fig.~\ref{fig:sub_bands_K2} we use $M_1=32$
 transmit antennas, equal power loading among the sub-bands and $P^{\rm tx}_{k,s}$= 22\,dBm (about 160\,mW) for each allocated sub-band and for each
 beam. That is, if the support of $g_k(f)$ is $N_k$ sub-bands,
 $k=1,2$, the total transmit power is $160(N_1+N_2)$\,mW.  We set $N_1=N_2=2$ in Fig.~\ref{fig:sub_bands_K2}(left),
 $N_1=4$ and $N_2=2$ in Fig.~\ref{fig:sub_bands_K2}(center) and
 $N_1=N_2=4$ in Fig.~\ref{fig:sub_bands_K2}(right).  In the figures
 the upper bound ``UB'' is computed using~\eqref{eq:upper_bound3}
 whereas ``Adapted beamforming'' technique is applied at the BS using
 the beamforming vectors in~\eqref{eq:adapted_bf}, where the angles
 $\beta_k$, $k=1,2$, are the directions of the center of the two
 physical IRSs as observed from the center of the BS array.
  We first observe that for a rank-2 system the upper bound ``UB'' is
  less tight than in the rank-1 case. Indeed, especially in the cases
  $N_1=N_2=2$ and $(N_1,N_2) =(4,2)$, it shows a rate about 50\% higher than
  that provided by the best technique, i.e., ``Max-eig phase''.  Also in this scenario, ``Max-eig phase''
  outperforms all other techniques, especially when the number of
  allocated sub-bands is small. Specifically, for $N_1=N_2=2$
  ``Max-eig phase'' provides 20\% rate increase w.r.t. ``NB Optimum'',
  which is its best competitor. Such rate increase reduces to about
  10\% in the case $N_1=4$, $N_2=2$, whereas it is almost negligible
  for $N_1=N_2=4$. Instead, the rate provided by ``NB central'' is
  almost linearly decreasing in all the considered cases.

\subsection{Summary of the results}
From the mathematical analysis introduced in Sections~\ref{sec:UB} and
from the simulation results shown, we can summarize the following messages,
which can be useful for the implementation of an IRS-aided
UWB network:
\begin{itemize}
\item although beam squint at both BS and IRS play an
  important role, the system performance is also strongly affected by
  the system geometry, as can be observed by Figs.~\ref{fig:spread1} and~\ref{fig:sub_bands_angle};
\item the upper bound proposed in~\eqref{eq:upper_bound3} is easy to
  compute {\color{black} and very tight, especially for single carrier ultra-wideband
  signals.  For signals with power density profiles such as in
  Figure~\ref{fig:Gf}(b) the upper-bound is always very tight for
  relative bandwidths up to about 15\% whereas the gap with ``Max-eig
  phase'' (the best performing technique) is noticeable for relative
  bandwiths greater than 15\% (see Figs. 5 and 7). This gap, however,
  does not exceed 30\% of the rate in the worst case, and thus
  represents a good approximation of the system performance.}
\item the ``NB central'' technique is the simplest to implement and,
  for sparse-spectrum signals,
  provides performance near to the optimum
  for signal bandwidth $B_w<30$\,GHz. However, its performance dramatically drops for
  $B_w>30$\,GHz. Instead, for UWB single-carrier signals, it performs
  always close to the upper bound for all the considered range of
  relative bandwidths;
\item the ``NB Optimum'' technique achieves, in general, good
  performance; 
\item the ``SCF'' algorithm, despite its complexity and its ability to
  provide a maximizer to~\eqref{eq:UCQP}, does not achieve superior
  performance in terms of rate, for it is most of the times worse or
  comparable to ``Max-eig phase'';
\item {\color{black} the SDR technique  performs worse than ``SCF'';}
\item the ``Max-eig phase'' solution seems to provide the best
  trade-off between performance and complexity for $B_w>30$\,GHz, large $L$, and
  a small number of allocated sub-bands, although, for very large $L$,
  its computational complexity might be unaffordable.
\end{itemize}

\section{Conclusions\label{sec:conclusions}}
We considered the problem of maximizing the achievable rate in a THz
communication system where both BS and IRS may generate the
beam-squint effect. We then proposed a set of techniques for
configuring the IRS in the presence of wideband signals and compared against
an upper bound. The channel model we used is realistic and
specific for THz communication, accounting for multipath due to
reflection from few large objects, molecular absorption, and
shadowing.  We specifically considered
two important scenarios envisaged for 6G THz communications: i) the
case of a single carrier UWB signal and ii) a multicarrier signal. 
We also provided an analytical condition granting the
optimality of the ``NB Central'' solution, which is the simplest to
implement.  
Our results pave the way for a fruitful application of IRSs and UWB
signaling to wireless networks, and indicate that the spectrum
allocation policy needs to take into account the system geometry and
the IRS configuration in order to achieve information-theoretical
performance.

\appendices 
\section{Proof of Propositions~\ref{prop:1} and~\ref{prop:1bis} \label{app:prop:1}}
The rate $R(\thetav)$ in~\eqref{eq:R} can be rewritten as
\begin{eqnarray}
  R(\thetav) &=& \int_{\Bc} \log \left|\Id+ \frac{1}{N_0}
  \Qm(f,\thetav)\right| \dd f \non
  &=& \int_{\Bc} \sum_{m=1}^{r_Q(f,\thetav)}\log
  \left(1+ \frac{\lambda_{Q,m}(f,\thetav)}{N_0}\right) \dd f
\end{eqnarray}
where the $M_2\times M_2$ matrix $\Qm(f,\thetav) = \Hm(f,\thetav)
\Gm(f)\Hm\Herm(f,\thetav)$ is positive definite, has rank
$r_Q(f,\thetav)=\rho(\Qm(f,\thetav))$ and has eigenvalues
$\lambda_{Q,m}(f,\thetav)$, $m=1,\ldots, r_Q(f,\thetav)$.
Given
the system parameters and for every frequency $f$, once $\thetav$ is
chosen, the eigenvalues $\lambda_{Q,m}(f,\thetav)$ are known, as well
as
\begin{equation}\label{eq:P_r_lambda}
  P^{\rm rx}(\thetav) = \int_{\Bc}\sum_{m=1}^{r_Q(f,\thetav)}\lambda_{Q,m}(f,\thetav)\dd f=\int_{\Bc}\trace\{\Qm(f,\thetav)\}\dd f
\end{equation}
which corresponds to the total received power.
Then, we can bound $R(\thetav)$ as follows:
\begin{eqnarray}
  R(\thetav)
  &\le& \max_{\yv(f)}\int_{\Bc}\sum_{m=1}^{r_Q(f)}\log\left(1 + \frac{y_{m}(f)}{N_0}\right) \dd f \label{eq:bound}
\end{eqnarray}
subject to $\int_{\Bc}\sum_{m=1}^{r_Q(f)} y_{m}(f)\dd f\mathord{=} P^{\rm rx}(\thetav)$,
$y_{m}(f)\mathord{\ge} 0$, $m=1,\ldots, r_Q(f)$, $\forall f\in\Bc$.
Here $\yv(f)=[y_1(f),\ldots, y_{M_2}(f)]\Tran$ is a vector of auxiliary
functions. 
Using~\eqref{eq:B_m}, the maximization over
$\yv(f)$ in~\eqref{eq:bound} can be solved by using the
Eulero-Lagrange formula.
We obtain
\begin{equation}
  R(\thetav) \le \sum_{m=1}^{M_2}B_m\log\left(1 + \frac{P^{\rm rx}(\thetav)}{N_0\sum_{m=1}^{M_2}B_m}\right)
\end{equation}
and, by consequence,~\eqref{eq:upper_bound}. 
As a special case, when the matrix $\Gm(f)$ is rank-1, it is clear that the matrix
$\Qm(f,\thetav)$ is also rank-1. Hence $B_m=0$ for
$m=2,\ldots,M_2$, and~\eqref{eq:upper_bound} reduces to~\eqref{eq:upper_bound_rank1}.
As a final step, in order to evaluate the r.h.s. of
~\eqref{eq:upper_bound} we need to solve the problem
$\max_{\thetav} P^{\rm rx}(\thetav)$ i.e., we look for the IRS configuration,
$\thetav$, maximizing the total received power.  By
recalling~\eqref{eq:P_r}, the definition of $\Qm(f,\thetav)$, and~\eqref{eq:H_f} we can
write 
\begin{eqnarray}
  P^{\rm rx}(\thetav)
&=& \int_{\Bc}\trace\left\{\Hm(f,\thetav) \Gm(f)\Hm\Herm(f,\thetav)\right\}\dd f \non
&=& \gammav\Herm \Tm \gammav +w +2\Re\left\{\qv\Herm \gammav\right\} \label{eq:Pr_gamma}
\end{eqnarray}
where $\gammav=[\gamma_1,\ldots,\gamma_L]\Tran$,
$\gamma_{\ell}=\ee^{\jj\theta_{\ell}}$, $\ell=1,\ldots,L$, $\Tm$ as in \eqref{eq:Tm}, and  $w$ and $\qv$ defined in~\eqref{eq:q_ell}, respectively.
It follows that
\begin{equation}
  \max_{\thetav}P^{\rm rx}(\thetav) = \max_{\gammav, |\gamma_{\ell}|=1, \forall \ell}\left(\gammav\Herm \Tm \gammav +w +2\Re\left\{\qv\Herm \gammav\right\}\right)\label{eq:UCQP_app}\,.
\end{equation}
Since $\Tm$ is positive definite and $\gammav\Herm\Tm\gammav$ is non-negative, an upper bound to~\eqref{eq:UCQP_app} can be obtained by relaxing
the $L$ unimodular constraints $|\gamma_{\ell}|=1$ to the
single quadratic constraint $\gammav\Herm\gammav=L$.
We get
\begin{eqnarray}
  && \hspace{-10ex}\max_{\gammav,  |\gamma_{\ell}|=1, \forall \ell}\left(\gammav\Herm \Tm \gammav\mathord{+}w\mathord{+}2\Re\left\{\qv\Herm \gammav\right\}\right) \non
  &\le&  \max_{\gammav,  \|\gammav\|^2=L} \gammav\Herm \Tm \gammav\mathord{+}w\mathord{+}2\max_{\gammav,  |\gamma_{\ell}|=1, \forall \ell}\Re\{\qv\Herm\gammav\} \label{eq:relax}\\
  &=& L\lambda_T^{\max} +w+2\sum_{q=1}^L|q_{\ell}|\label{eq:relax2}\
\end{eqnarray}
where $\lambda_T^{\max}$ is the largest eigenvalue of $\Tm$.
A computable expression for $r_Q(f)$, the upper bound to $r_Q(f,
\thetav)$, for the model in \eqref{eq:H_f}-\eqref{eq:Hl_f} can be
obtained as follows. The rank $r_Q(f,\thetav)$ can be first
upper-bounded by
\begin{eqnarray}
  r_Q(f,\thetav) 
  &=& \rho( \Hm(f,\thetav)\Gm(f)\Hm\Herm(f,\thetav)) \non
  &\le& \min\{ \rho(\Hm(f,\thetav)),\rho(\Gm(f))\}\,.
\end{eqnarray}
Next, we recall%
~\eqref{eq:H_f} and~\eqref{eq:Hl_f} and observe that
\[ \rho(\Hm(f,\thetav))\le \min\left\{M_2,\rho\left( \sum_{\ell=1}^L \ee^{\jj \theta_\ell} \Hm_{\ell}(f)\right) +\rho(\bar{\Hm}(f))\right\}\,. \]
Since $\Hm_{\ell}(f)$ is a rank-1 matrix, $\forall \ell=1,\ldots,L$, we can write  $\Hm_{\ell}(f)=\vv_{\ell}(f)\wv_{\ell}(f)\Tran$
for certain vectors $\vv_{\ell}(f)$ and $\wv_{\ell}(f)$ and, thus, 
$\sum_{\ell=1}^L \ee^{\jj \theta_\ell} \Hm_{\ell}(f) = \Vm(f)\Thetam\Wm(f)\Tran$
where $\Vm(f) = [\vv_1(f),\ldots, \vv_L(f)]$, $\Wm(f) =
[\wv_1(f),\ldots, \wv_L(f)]$ and $\Thetam$ is a diagonal matrix whose
entries are $ \ee^{\jj \theta_\ell}$, $\ell=1,\ldots,L$. Since
$\Thetam$ has rank $L$ (it is diagonal with non-zero diagonal
elements) we have $\rho(\Vm(f)\Thetam\Wm(f)\Tran)\le \min\{L,\rho(\Vm(f)),\rho(\Wm(f))\}$. Then, we can upper-bound $r_Q(f,\thetav)$ as~\eqref{eq:rank_bound}.

\section{Proof of Proposition~\ref{prop:2}\label{app:prop:2}}
The power spectral density matrix for a monochromatic signal at
frequency $f_0$ can be written as $\Gm(f) =\Gm_0 \delta(f-f_0)$. By
using~\eqref{eq:hyp1}, the entry $(\ell, \ell')$ of matrix $\Tm$ is
given by
\begin{eqnarray} 
  [\Tm]_{\ell,\ell'}
  &=& \int_{\Bc} \trace\left\{\Hm_{\ell}(f) \Gm(f)\Hm_{\ell'}\Herm(f) \right\} \dd f \non
  &=&K_{\ell}K_{\ell'}\trace\left\{\Am(f_0) \Gm_0\Am\Herm(f_0) \right\} \ee^{\jj 2\pi f_0 (\tau_{\ell'} - \tau_{\ell})}\,. \nonumber
\end{eqnarray}
Hence $\Tm = M_0\zv\zv\Herm$ where $M_0=\trace\left\{\Am(f_0)\Gm_0\Am\Herm(f_0) \right\}$, $\zv=[z_1,\ldots,z_L]\Tran$ and
$z_{\ell}=K_{\ell}\ee^{-\jj 2\pi f_0 \tau_{\ell}}$, $\ell=1,\ldots,L$.
Since the coefficients $K_{\ell}$ are positive, the term
$\gammav\Herm\Tm\gammav$ is
given by
\begin{eqnarray}
  \gammav\Herm\Tm\gammav
  &=& M_0\left|\sum_{\ell=1}^L z_{\ell}^*\gamma_{\ell} \right|^2
  \le  
   M_0\sum_{\ell=1}^L K_{\ell}^2,
\end{eqnarray}
where the equality is obtained for $\gammav =\gammav^{\rm NB}(f_0)$
in~\eqref{eq:NB}.  Therefore $\gammav^{\rm NB}(f_0)$ is the maximizer
of $\gammav\Herm\Tm\gammav$ and, hence, the optimal solution for a
monochromatic signal.

\section{Proof of Proposition~\ref{prop:3}\label{app:prop:3}}
Let us compute entry $(\ell, \ell')$ of matrix $\Tm$ when \eqref{eq:hyp1} holds. We have:
\begin{eqnarray}
[\Tm]_{\ell,\ell'}  
 &=& \int_{\Bc} \trace\left\{\Hm_{\ell}(f) \Gm(f) \Hm_{\ell'}\Herm(f) \right\} \dd f \non
&=&  K_{\ell}K_{\ell'}\int_{\Bc} \trace\left\{\Cm(f) \right\} \ee^{\jj 2\pi f (\tau_{\ell'} - \tau_{\ell})} \dd f \non
  &=&K_{\ell}K_{\ell'} \ee^{\jj 2\pi f_0 (\tau_{\ell'}- \tau_{\ell})} \non
      && \cdot\int_{\mathord{-}\frac{B_w}{2}}^{\frac{B_w}{2}}\trace\left\{\Cm(f\mathord{+}f_0)\right\}\cdot \ee^{j2\pi f (\tau_{\ell'}\mathord{-}\tau_{\ell})} \dd f\,.\nonumber
\end{eqnarray}
where $\Cm(f)\triangleq \Am(f) \Gm(f) \Am\Herm(f)$.
Now, since the received signal spectrum is even around $f_0$, the
integral is real. Moreover, because of the assumption on the positive
semi-definiteness of $\Gm(f)$, made in Section~\ref{sec:model}, the trace in the integral
is non-negative. Defining this trace as $M(f)$, we get
\begin{equation}
  [\Tm]_{\ell,\ell'} \mathord{=} \ee^{\jj 2\pi f_0 (\tau_{\ell'} \mathord{-} \tau_{\ell})} \int_{-\frac{B_w}{2}}^{\frac{B_w}{2}}\hspace{-2ex}K_{\ell}K_{\ell'} M(f) \cos \left( 2\pi f (\tau_{\ell'} \mathord{-}\tau_{\ell})\right) \dd f\,.
\end{equation}
If condition \eqref{eq:f_cond} is satisfied, the integrand is positive
for every $f$, so that the integral is positive. Thus, we have
\begin{equation}
\Tm = \widetilde{\Tm} \odot \gammav^{\rm NB}(f_0) \left(\gammav^{\rm
    NB}(f_0)\right)\Herm
\end{equation}
where $\widetilde{\Tm}$ is a matrix with
all entries real positive. The solution of~\eqref{eq:UCQP_no_mpath} with such a matrix
is well known to be $\gammav^{\rm NB}(f_0)$~\cite{Stoica14}.

\section{Proof of Proposition~\ref{prop:4}\label{app:prop:4}}
  We first observe that the solution of~\eqref{eq:UCQP_no_mpath} does not
  change if we multiply the matrix $\Tm$ by a
  constant, say $1/L$. So, when \eqref{eq:case2} is satisfied, we can write $\frac1{L}\Tm = \Zm \Mm \Zm\Herm$
  where $\Mm{=}\diag(M_1, \dots, M_S)$ and $\Zm{=}[\zv_1/\sqrt{L},\dots, \zv_S/\sqrt{L}]$.  When $L \rightarrow \infty$, it
  is easy to see that $\frac{1}{L}\zv_s\Herm \zv_{s'} \rightarrow \delta_{s,s'}$, where $\delta_{s,s'}$ is the Kronecker $\delta$ function, unless $(f_s - f_{s'}) \tau$ is semi-integer, a case which we
  exclude since it has vanishing probability as $L\to \infty$. Thus,
  for $L \rightarrow \infty$, $\Zm \Mm \Zm\Herm$  becomes the
  eigenvalue decomposition of $\Tm$ and $M_1, \dots, M_S$
  its non-zero eigenvalues. Correspondingly,
  $\gammav^{\rm NB}(f_{s^*})$ becomes the eigenvector associated
  with the largest eigenvalue, yielding \beq \label{eq:smart} P^{\rm rx}(\gammav^{\mathrm{NB}}(f_{s^*}))=
  (\gammav^{\mathrm{NB}}(f_{s^*})) \Herm \Tm
  \gammav^{\mathrm{NB}}(f_{s^*}) 
  \eeq 
  which is clearly the largest
  possible value for $P^{\rm rx}$.
  
\bibliographystyle{IEEEtran2}
\bibliography{refs}

\begin{thebibliography}{10}
\providecommand{\url}[1]{#1}
\csname url@samestyle\endcsname
\providecommand{\newblock}{\relax}
\providecommand{\bibinfo}[2]{#2}
\providecommand{\BIBentrySTDinterwordspacing}{\spaceskip=0pt\relax}
\providecommand{\BIBentryALTinterwordstretchfactor}{4}
\providecommand{\BIBentryALTinterwordspacing}{\spaceskip=\fontdimen2\font plus
\BIBentryALTinterwordstretchfactor\fontdimen3\font minus
  \fontdimen4\font\relax}
\providecommand{\BIBforeignlanguage}[2]{{%
\expandafter\ifx\csname l@#1\endcsname\relax
\typeout{** WARNING: IEEEtran.bst: No hyphenation pattern has been}%
\typeout{** loaded for the language `#1'. Using the pattern for}%
\typeout{** the default language instead.}%
\else
\language=\csname l@#1\endcsname
\fi
#2}}
\providecommand{\BIBdecl}{\relax}
\BIBdecl

\bibitem{noiICC2023}
A.~Nordio, L.~Dossi, A.~Tarable, and G.~Virone, ``Near-field {IRS}
  configuration techniques for wideband signals and {THz} communications,'' in
  \emph{2023 IEEE International Conference on Communications Workshops (ICC
  Workshops)}, 2023. doi: 10.1109/ICCWorkshops57953.2023.10283765 pp.
  1198--1203.

\bibitem{Huq2019}
K.~M.~S. Huq, S.~A. Busari, J.~Rodriguez, V.~Frascolla, W.~Bazzi, and D.~C.
  Sicker, ``Terahertz-enabled wireless system for beyond-{5G} ultra-fast
  networks: A brief survey,'' \emph{IEEE Network}, vol.~33, no.~4, pp. 89--95,
  2019.

\bibitem{DRL-beamforming}
I.~Ahmed, K.~Shahid, and H.~Khammari, ``{DRL} based beam selection and hybrid
  beamforming for intelligent reflective surface assisted massive {MIMO}
  system,'' in \emph{2023 IEEE 97th Vehicular Technology Conference
  (VTC2023-Spring)}, 2023, pp. 1--6.

\bibitem{DiRenzo}
M.~Di~Renzo \emph{et~al.}, ``Smart radio environments empowered by
  reconfigurable {AI} meta-surfaces: an idea whose time has come,''
  \emph{EURASIP Journal Wireless Communication Networks}, vol. 129, May 2019.

\bibitem{HMIMO}
T.~Gong, P.~Gavriilidis, R.~Ji, C.~Huang, G.~C. Alexandropoulos, L.~Wei,
  Z.~Zhang, M.~Debbah, H.~V. Poor, and C.~Yuen, ``Holographic {MIMO}
  communications: Theoretical foundations, enabling technologies, and future
  directions,'' \emph{IEEE Communications Surveys \& Tutorials}, pp. 1--1,
  2023.

\bibitem{SpatialWideband}
B.~Wang, F.~Gao, S.~Jin, H.~Lin, G.~Y. Li, S.~Sun, and T.~S. Rappaport,
  ``Spatial-wideband effect in massive {MIMO} with application in mmwave
  systems,'' \emph{IEEE Communications Magazine}, vol.~56, no.~12, pp.
  134--141, 2018.

\bibitem{Delay_phase_precoding2022}
L.~Dai, J.~Tan, Z.~Chen, and H.~V. Poor, ``Delay-phase precoding for wideband
  {THz} massive {MIMO},'' \emph{IEEE Transactions on Wireless Communications},
  vol.~21, no.~9, pp. 7271--7286, 2022.

\bibitem{WB_beamforming_mMIMO}
F.~Gao, B.~Wang, C.~Xing, J.~An, and G.~Y. Li, ``Wideband beamforming for
  hybrid massive {MIMO} terahertz communications,'' \emph{IEEE Journal on
  Selected Areas in Communications}, vol.~39, no.~6, pp. 1725--1740, 2021.

\bibitem{Behrooz2023}
R.~Wang, Y.~Yang, B.~Makki, and A.~Shamim, ``A wideband reconfigurable
  intelligent surface for {5G} millimeter-wave applications,'' 2023. doi:
  10.48550/arXiv.2304.11572

\bibitem{survey}
S.~Gong, X.~Lu, D.~T. Hoang, D.~Niyato, L.~Shu, D.~I. Kim, and Y.-C. Liang,
  ``Toward smart wireless communications via intelligent reflecting surfaces: A
  contemporary survey,'' \emph{IEEE Communications Surveys \& Tutorials},
  vol.~22, no.~4, pp. 2283--2314, Apr. 2020.

\bibitem{wideband_IRSmodel}
W.~Cai, R.~Liu, Y.~Liu, M.~Li, and Q.~Liu, ``Practical modeling and beamforming
  for intelligent reflecting surface aided wideband systems,'' \emph{IEEE
  Communications Letters}, vol.~24, no.~7, pp. 1568--1571, 2020.

\bibitem{Zeng2022}
S.~Zeng, H.~Zhang, B.~Di, Y.~Liu, M.~D. Renzo, Z.~Han, H.~V. Poor, and L.~Song,
  ``Intelligent omni-surfaces: Reflection-refraction circuit model,
  full-dimensional beamforming, and system implementation,'' \emph{IEEE
  Transactions on Comm.}, vol.~70, no.~11, pp. 7711--7727, 2022.

\bibitem{BeamSquintMitigating}
Y.~Chen, D.~Chen, and T.~Jiang, ``Beam-squint mitigating in reconfigurable
  intelligent surface aided wideband mmwave communications,'' in \emph{2021
  IEEE Wireless Communications and Networking Conference (WCNC)}, 2021, pp.
  1--6.

\bibitem{wideband_THz}
\BIBentryALTinterwordspacing
D.~Konstantinos, s.~D.~A. Styliano, Q.~N. Hien, B.~Boris, and M.~Michail,
  ``Intelligent reflecting surface-aided wideband {THz} communications:
  Modeling and analysis,'' \emph{IEEE WCNC}, 2021. [Online]. Available:
  \url{https://https://arxiv.org/abs/2110.15768}
\BIBentrySTDinterwordspacing

\bibitem{rainbow}
M.~Cui, L.~Dai, Z.~Wang, S.~Zhou, and N.~Ge, ``Near-field rainbow: Wideband
  beam training for {XL-MIMO},'' \emph{IEEE Transactions on Wireless
  Communications}, pp. 1--1, 2022.

\bibitem{UWB_THz_IRS}
W.~Hao, F.~Zhou, M.~Zeng, O.~A. Dobre, and N.~Al-Dhahir, ``Ultra wideband {THz
  IRS} communications: Applications, challenges, key techniques, and research
  opportunities,'' \emph{IEEE Network}, vol.~36, no.~6, pp. 214--220, 2022.

\bibitem{WB_estimation}
S.~Ma, W.~Shen, J.~An, and L.~Hanzo, ``Wideband channel estimation for
  {IRS}-aided systems in the face of beam squint,'' \emph{IEEE Transactions on
  Wireless Communications}, vol.~20, no.~10, pp. 6240--6253, 2021.

\bibitem{Hashemi2008}
H.~Hashemi, T.-s. Chu, and J.~Roderick, ``Integrated true-time-delay-based
  ultra-wideband array processing,'' \emph{IEEE Communications Magazine},
  vol.~46, no.~9, pp. 162--172, 2008.

\bibitem{DelayAdjustable_Hanzo}
\BIBentryALTinterwordspacing
J.~An, C.~Xu, D.~W.~K. Ng, C.~Yuen, L.~Gan, and L.~Hanzo, ``Reconfigurable
  intelligent surface-enhanced {OFDM} communications via delay adjustable
  metasurface,'' \emph{CoRR}, vol. abs/2110.09291, 2021. [Online]. Available:
  \url{https://arxiv.org/abs/2110.09291}
\BIBentrySTDinterwordspacing

\bibitem{prototype}
R.~Fara, P.~Ratajczak, D.-T. Phan-Huy, A.~Ourir, M.~Di~Renzo, and J.~de~Rosny,
  ``A prototype of reconfigurable intelligent surface with continuous control
  of the reflection phase,'' \emph{IEEE Wireless Communications}, vol.~29,
  no.~1, pp. 70--77, 2022.

\bibitem{noiTWC}
A.~Tarable, F.~Malandrino, L.~Dossi, R.~Nebuloni, G.~Virone, and A.~Nordio,
  ``Optimization of {IRS}-aided sub-{THz} communications under practical design
  constraints,'' \emph{IEEE Transactions on Wireless Communications}, vol.~21,
  no.~12, pp. 10\,824--10\,838, 2022.

\bibitem{Molecular}
J.~Kokkoniemi, J.~Lehtomäki, and M.~Juntti, ``Simple molecular absorption loss
  model for 200–450 gigahertz frequency band,'' in \emph{2019 European
  Conference on Networks and Communications (EuCNC)}, 2019. doi:
  10.1109/EuCNC.2019.8801950 pp. 219--223.

\bibitem{propagation}
C.~Han and Y.~Chen, ``Propagation modeling for wireless communications in the
  terahertz band,'' \emph{IEEE Communications Magazine}, vol.~56, no.~6, pp.
  96--101, 2018.

\bibitem{Pan_2020}
C.~Pan, H.~Ren, K.~Wang, M.~Elkashlan, A.~Nallanathan, J.~Wang, and L.~Hanzo,
  ``Intelligent reflecting surface aided mimo broadcasting for simultaneous
  wireless information and power transfer,'' \emph{IEEE Journal on Selected
  Areas in Communications}, vol.~38, no.~8, pp. 1719--1734, 2020.

\bibitem{Stoica14}
M.~Soltanalian and P.~Stoica, ``Designing unimodular codes via quadratic
  optimization,'' \emph{IEEE Transactions on Signal Processing}, vol.~62,
  no.~5, pp. 1221--1234, 2014.

\bibitem{Aldayel17}
O.~Aldayel, V.~Monga, and M.~Rangaswamy, ``Tractable transmit {MIMO}
  beampattern design under a constant modulus constraint,'' \emph{IEEE
  Transactions on Signal Processing}, vol.~65, no.~10, pp. 2588--2599, 2017.

\bibitem{Larsson_2020}
O.~{\"O}zdogan, E.~Bj{\"o}rnson, and E.~G. Larsson, ``Intelligent reflecting
  surfaces: Physics, propagation, and pathloss modeling,'' \emph{IEEE Wireless
  Communications Letters}, vol.~9, no.~5, pp. 581--585, May 2020.

\bibitem{3gppchanmodel}
3GPP, ``{5G; Study on channel model for frequencies from 0.5 to 100 GHz -
  Release 14},'' {3rd Generation Partnership Project (3GPP)}, Tech. Rep.
  38.901, 2017.

\bibitem{ITU-Attenuation}
ITU-R, ``Attenuation by atmospheric gases and related effects,'' International
  Telecommunication Union, Recommendation p.676-13, 2022.

\bibitem{reflection}
ITU-T, ``Effects of building materials and structures on radiowave propagation
  above about 100 {MHz},'' International Telecommunication Union,
  Recommendation P.2060-1, 2015.

\bibitem{Shafie2019}
A.~Shafie, N.~Yang, C.~Han, J.~M. Jornet, M.~Juntti, and T.~Kurner, ``Terahertz
  communications for {6G} and beyond wireless networks: Challenges, key
  advancements, and opportunities,'' \emph{IEEE Network}, pp. 1--8, 2022.

\bibitem{SemidefiniteRelaxation}
Z.-q. Luo, W.-k. Ma, A.~M.-c. So, Y.~Ye, and S.~Zhang, ``Semidefinite
  relaxation of quadratic optimization problems,'' \emph{IEEE Signal Processing
  Magazine}, vol.~27, no.~3, pp. 20--34, 2010.

\bibitem{Wu-Zhang2018}
Q.~Wu and R.~Zhang, ``Intelligent reflecting surface enhanced wireless network:
  Joint active and passive beamforming design,'' in \emph{2018 IEEE Global
  Communications Conference (GLOBECOM)}, 2018. doi: 10.1109/GLOCOM.2018.8647620
  pp. 1--6.

\end{thebibliography}

\end{document}